\documentclass[sigconf]{acmart}

\usepackage{booktabs} 

\renewcommand\footnotetextcopyrightpermission[1]{} 
\begin{document}
\title{A Qualitative Post-Experience Method for Evaluating Changes in VR Presence Experience Over Time}


\author{Christian Mai}

\affiliation{%
  \institution{LMU Munich}
  \postcode{43017-6221}
}
\email{christian.mai@ifi.lmu.de}

\author{Heinrich Hu\ss{}mann}

\affiliation{
  \institution{LMU Munich}}
\email{hussmann@ifi.lmu.de}

\renewcommand{\shortauthors}{C. Mai \& H. Hu\ss{}mann}

%


\begin{abstract}
A particular measure to evaluate an head-mounted display (HMD) based experience is the state of feeling present in virtual reality.
Interruptions of a presence experience -- break in presence (BIP) -- appearing over time, need to be detected to assess and improve an application.
Existing methods either lack in taking these BIPs into account -- questionnaires -- or are complex in their application and evaluation -- physiological and behavioral measures --.
To provide a practical approach, we propose a post-experience method in which the users reflect on their experience by drawing a line, indicating their experienced state of presence, in a paper-based drawing template.
The amplitude of the drawn line represents the variation of their presence experience over time.
We propose a descriptive model that describes temporal variations in the drawings by the definition of relevant points over time -- e.g., putting on the HMD --, phases of the experience -- e.g., transition into VR -- and parameters --e.g., the transition time--.
The descriptive model enables us to objectively evaluate user drawings and represent the course of the drawings by a defined set of paramters.
An exploratory user study (N=30) showed that the drawings are very consistent, the method can detect all BIPs and shows good indications for representing the intensity of a BIP.
With our method practitioners and researchers can accelerate the evaluation and optimization of experiences by evaluating BIPs. 
The possibilty to store objective paramters paves the way for automized evaluation methods and big data approaches.
\end{abstract}


\begin{teaserfigure}
\centering
  \includegraphics[width=\textwidth]{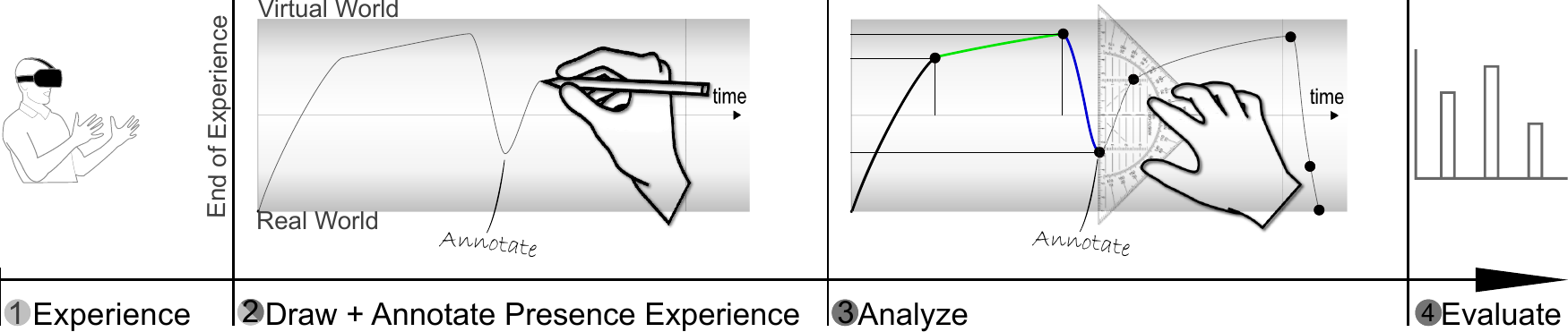}
  \captionof{figure}{We propose a post-experience drawing method to collect data about the temporal cause of a presence experience. After ending the experience (1), the user is asked to \emph{draw} the course of the presence experience (2) into a template we provide. The horizontal axis represents the time, the vertical axis the experienced state of presence, either being in the real or in the virtual world. The user is asked to \emph{annotate} the drawing (2), in particular breaks in presence and phases of constant presence, to relate the drawing to events during the experience. A supervising person might support the annotation. Afterwards the drawing is \emph{analyzed} by identifying distinctive points and phases by an expert (3). We propose a \emph{descriptive model} that explains these distinctive points and phases based on related work on the presence phenomenon. The points' position is stored as compact numerical data in a database, for storage, comparison and \emph{evaluation} (4) purposes.}
  \label{fig:teaser}
\end{teaserfigure}

\maketitle

\section{Introduction}


The availability of consumer grade head-mounted displays (HMD) fosters the usage by a growing group of users in the research community and gives industrial users new commercial perspectives. 
To improve VR experiences during development and to answer research questions, methods and tools are needed to evaluate these experiences. 
A unique measure used for VR systems is the quantification of the user's state to feel present in the VR. 

The state of feeling present in the VR was shown to vary over time \cite{Garau.2008,Knibbe:2018:DCE:3173574.3174057,Slater:2000,sproll13,4811024}.
Temporal variations can have short-term as well as long-term effects on the user.
Particularly negative effects in the course of an experience are noticed by the user, so-called breaks in presence (BIPs) \cite{Slater:2000}.
An example are people acting in the real world touching an HMD user.
Also cognitive processes of the user, like getting bored or frustrated in the VR experience, can make a user attention shift from the virtual environment back to the real world \cite{sproll13}.
Practical tools are missing that asses the presence experience or to detect these disturbances.



Today more than 90\% of studies using the state of presence as a measure rely on questionnaires \cite{cummings_how_2015,mai2018,rosa.14}.
Questionnaires are typically used post experience which can distort results due to inaccurate recall and are not able to detect variations of the presence over time \cite{Garau.2008,Liebold2017}.
Interviews, in addition to questionnaires, can support the evaluation of temporal events.
However, interviews do not give a continuous view on the experience, require interpretation and effort by the examiner and do not give a quantification of the state of presence.
Physiological measurements enable a continuous evaluation of the users' state of being present.
They are complicated to use, intrusive to a user and the association of the measured signal and the state of presence is often not clear \cite{Garau.2008,Liebold2017}.
Behavioral observations are not intrusive to a user.
In most cases, behavioral observations do not enable continuous evaluation as they need to follow a particular protocol.
This protocol might be only suitable to detect the reaction ones or only to specific points in time -- e.g., an unexpected car approaching the user --~\cite{Liebold2017}. 

Therefore the \textbf{main goal} of this paper is to provide a method that enables quick, simple to use and reliable evaluation of the presence experience considering temporal variations, in particular, the detection of BIPs.
To achieve this we propose a drawing method based on the work of Garau~\cite{Garau.2008}.
The drawing method supports HMD users' reflection on the VR experience, by drawing a continuous line along a time axis (Figure~\ref{fig:teaser}).
The amplitude of the drawn line expresses the variations of the presence state during the cause of the experience, indicating the experience to be in a physical room or a virtual environment.
As Garau's approach is in a conceptual state it does not yet describe a full method and does not incorporate related work on temporal variations of the presence state.
Hence we designed a drawing template that takes into account related work and give recommendations for its application.
In an iterative design process, we optimized the template by taking into account related work and user needs while drawing.
To enable quantitative evaluation and storage of the data in a database, we propose a descriptive model (Figure \ref{fig:drawingarea}) that we derive from related work and experience from application.
It replaces the users' drawings by describing the important points and phases that occur during a VR experience, e.g., the point describing the end of the transition into VR.R (e.g.,~\cite{4811024}).
From the formal description we then derive concrete phases -- e.g., transitioning into VR--, and further parameters describing temporal effects and changes of the presence state.


In an application example with 30 participants naive to the method, we demonstrate the feasibility and potential of the proposed drawing method.
In the progress of a 15 minutes long experience, they were exposed to five breaks in presence with varying intensity.
They did not know about the breaks in advance, nor where they asked to identify breaks during the experience.
The resulting drawings are valid and can be used for the evaluation.
We found that our method is able to detect 100\% of the breaks.
The detection rate will remain 100\% with less participants.
Further we were able to find disturbances that we did not detect with traditional usability methods during the design process of the study.
Additionally there is strong evidence that the drawings reflect the strength of a break.


With our work we contribute:
\begin{itemize}
    \item A drawing template and method to collect user reports about their presence experience over time.
    \item A model describing temporal variations of presence using parameters to replace the drawings by compact numerical data for storage, comparison and evaluation purposes.
    \item Insights in the application of the drawing method.
\end{itemize}

\section{Background}
As the term presence is crucial to our work, we will introduce it first.
We continue by presenting literature that assesses and conceptualizes temporal variations of the presence experience.


\subsection{Presence}
There are multiple definitions of the term \textit{presence} (e.g.~\cite{minsky_telepresence_1980,schubert_experience_2001,Slater97,Witmer98}).
We distinguish presence from the term immersion.
Immersion describes the measurable attributes of a system enabling to evoke the feeling of being present.
The sense of presence is a cognitive construct that integrates incoming stimuli with user attributes, leading to a variable state of presence.
The term presence is often referred to a feeling of ''being there``, the spatial presence~\cite{Schubert2003} or place illusion~\cite{Slater3549}.
Other forms of presence can be experienced like involvement~\cite{Schubert2003} in a story or the general plausibility~\cite{Slater3549}/ realness~\cite{Schubert2003} of an experience and social presence~\cite{heeter92}.
Our work is not particularly linked to one of the definitions, as the usage of the drawing method depends on the question one asks the users.
The supervising person of the proposed drawing method needs to consider which definition s/he wants to follow and ask the participant the according questions.
In our user study, we follow the definition of Schubert and colleagues \cite{schubert_experience_2001}.
They propose the three components realness, involvement and spatial presence that influence the presence feeling.

\subsection{Assessing Temporal Variations of Presence}
We focus on user self-reports.
To date, self-reports offer the most practical way to evaluate an HMD-based experience, which is reflected in a high proportion of the use of the presence evaluation method~\cite{cummings_how_2015,mai2018,rosa.14}.
Continuously asking the users about their presence state during the experience is the simplest form of user self-reports.
As this would foster a reorientation of a user's attention to the real world, Slater~\cite{Slater:2000} introduced the concept of a \emph{virtual presence counter}. 

The presence counter asks for the experience of breaks in presence when they appear, a binary measure, instead of reporting a presence state on a scale.
Slater argues that users are not aware of the transition to VR and therefore cannot reflect the presence status in which they find themselves. 
Users can very well explain when they transition back into the real world, as a result from a break in presence (BIP). 
Although the method is convenient, it might overemphasize the report of weak BIPs. 
The breaks might have been unrecognized when left out for further attention, and the method is not feasible to determine the strength of a BIP~\cite{Liebold2017}.

Other work used mechanical sliders to give the user the opportunity to report the feeling of being present by manipulating the position of a mechanical or virtual slider over time~\cite{chung12,Ijsselsteijn98}. 
The physicality of the sliders constantly reminds the users about the real world, although the authors tried to limit this effect, and therefore negatively influence the feeling of being present \cite{chung12,Ijsselsteijn98}. 
The users' ability to interact in the VR is limited, as they continuously need one hand to manipulate the slider.
To avoid interruptions during the experience, we aim for a post-experience measure, which adds the possibility to evaluate temporal experiences in particular with the goal to detect and rate BIPs.

Besides the collection of data, there are a few advances to conceptualize the temporal variations of presence.
Slater's presence counter gives a stochastic view on the appearance of breaks, but does not define the temporal variation in more detail~\cite{Slater:2000}.
Liebold and colleagues discuss the rationale of the strength of break in presence, based on human cognition \cite{Liebold2017}. 
They provide parameters to rate the degree of disruption created by a BIP for the HMD user.
Their detection method relies on a physiological measure and does not give a formal description of temporal variations.
Chung and colleagues describe the temporal course of a break in presence by the impact length and the recovery time from a BIP \cite{Chung:2009:MTV:1670252.1670287}.
The impact length described by them is the drop in the presence value to the lowest value during the BIP.
They do not define the parameters describing the points chosen for their measurements.
They do not report how they handle plateaus in the drawings, next to or during the BIP.
Garau and colleagues~\cite{Garau.2008} introduced a drawing method to gain insights on temporal variations during an HMD based experience.
They found that the reflection on the experience of drawing supported the users' memory of the course of temporal fluctuations.
The drawing template was not grounded in related work and therefore did not give any guidance to the user when drawing the experience.
Therefore the drawings were very diverse and did not show phases of transitioning into or out of VR.
As the drawings were thought to support the discussion, the presented analysis sticks to a mere description of the drawings.

Kujala and colleagues~\cite{Kujala:2011:UCM:2051372.2051531}, provide a method to assess user experience over several month and was positivly confirmed in other studies~\cite{Varsaluoma:2014:MRU:2676467.2676485,Sahar:2014:CEE:2676467.2676500}. 
In their retrospective method, the user has to draw the degree of pleasantness of using, e.g., a smartphone, being either positive or negative.
They show that the method is beneficial as it supports the users' remembering of using a phone over several month.
The participants were asked to leave comments on the curve, which revealed that in the long term experience especially the hedonic quality of the system was remembered~\cite{Kujala:2011:UCM:2051372.2051531}, which was confirmed by Sahara and colleagues~\cite{Sahar:2014:CEE:2676467.2676500}.
The proposed drawing method does not fit the context of a VR experience that last only several minutes or hours and is assessed immediately afterwards.
Therefore the results on the application of the drawing method can not be transferred.
An exemplary difference would be a better remembrance of pragmatic qualities of the system, as the time between experience and evaluation is much shorter.
In addition, the question about the presence experience is different from the question of the pleasantness of a product.
Different cognitive models by the participants might need a different design of the drawing template and influence the way users draw the line.


We conclude from the related work that there is a need and several attempts to asses temporal effects of a presence experience.
Letting the participants draw their presence experience is a feasible approach to the problem, however a feasible method is missing.
Further related work does not provide a complete model to describe the course of the temporal fluctuations.
Hence we propose a descriptive model that links existing work together and provides a model of temporal fluctuations.
We use the model primarily to analyze the outcome of our drawing method, but it can also be used as a communication model to discuss temporal effects of a VR experience on as it summarizes all possible effects.

\section{Drawing the Presence Experience}
\label{sec:drawingmethod}
\begin{figure}[b]
    \centering
    \includegraphics[width=\columnwidth]{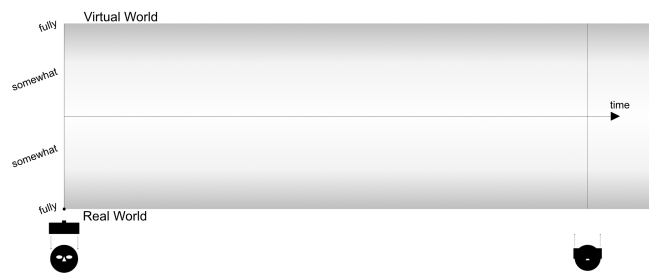}
    \caption {The basic drawing template can be used to collect user reports of temporal variations in the presence experience after having an HMD experience. The left icon indicates the moment the user put on the HMD, the right icon the moment the user took of the HMD. The basic template should be accompanied by a definition of the term presence (e.g., from the IPQ~\cite{Schubert2003}), according to the preference of the conducting person.}
    \label{fig:drawingtemplate}
\end{figure}

We want to provide a method that is feasible to detect and evaluater the temporal course of presence in a VR experience. 
Therefore we propose a post-experience drawing method that supports the HMD user in expressing and reflecting the experience.

\subsection{Design Process of the Drawing Template}
To theoretical motivations we add \emph{practical findings} from a collection of 60 presence drawings of four different user studies, unrelated to the study presented in this work.
We collected these drawings to better understand and support participants usage of the drawing template.
The participants were presented variations of the drawing template that were improved between sessions during an iterative design process.
Participants were asked to draw their feeling of spatial allocation in the VR or the real world, the experienced involvement in the experience and the general plausibility of the experience.
The collection of drawings was accompanied by interviews with the participants about the course of drawing and their understanding of the drawing template.
The Template for the drawings included event ticks for particular tasks in a linear story-line along the time axis for about one third of the participants (similar to Figure \ref{fig:Transitionalphases}).
The other two thirds were not provided event ticks as the experience was exploratory without an order of events (Figure \ref{fig:drawingtemplate}).

Figure \ref{fig:drawingtemplate} shows the final version of the template.
As a result from the iterative design process, we added a dot that indicates the starting point for the drawing.
We introduce linear, gradual transition from pure white to 25\% grey from the time axis towards the extremes (See Figure \ref{fig:drawingtemplate}).
By this, we give the user orientation while drawing along the time axis and indicate the states in which the other world is forgotten.

\subsection{The Drawing Template}
Previous work uses a drawing area with a horizontal time axis and the presence state on top of it, without negative values \cite{Chung:2009:MTV:1670252.1670287,Garau.2008,IJsselsteijn00}.
Therefore, the drawing area is limited to the bottom, which indicates to be fully present in the real world and open at the top.

\begin{figure*}
    \centering
    \includegraphics{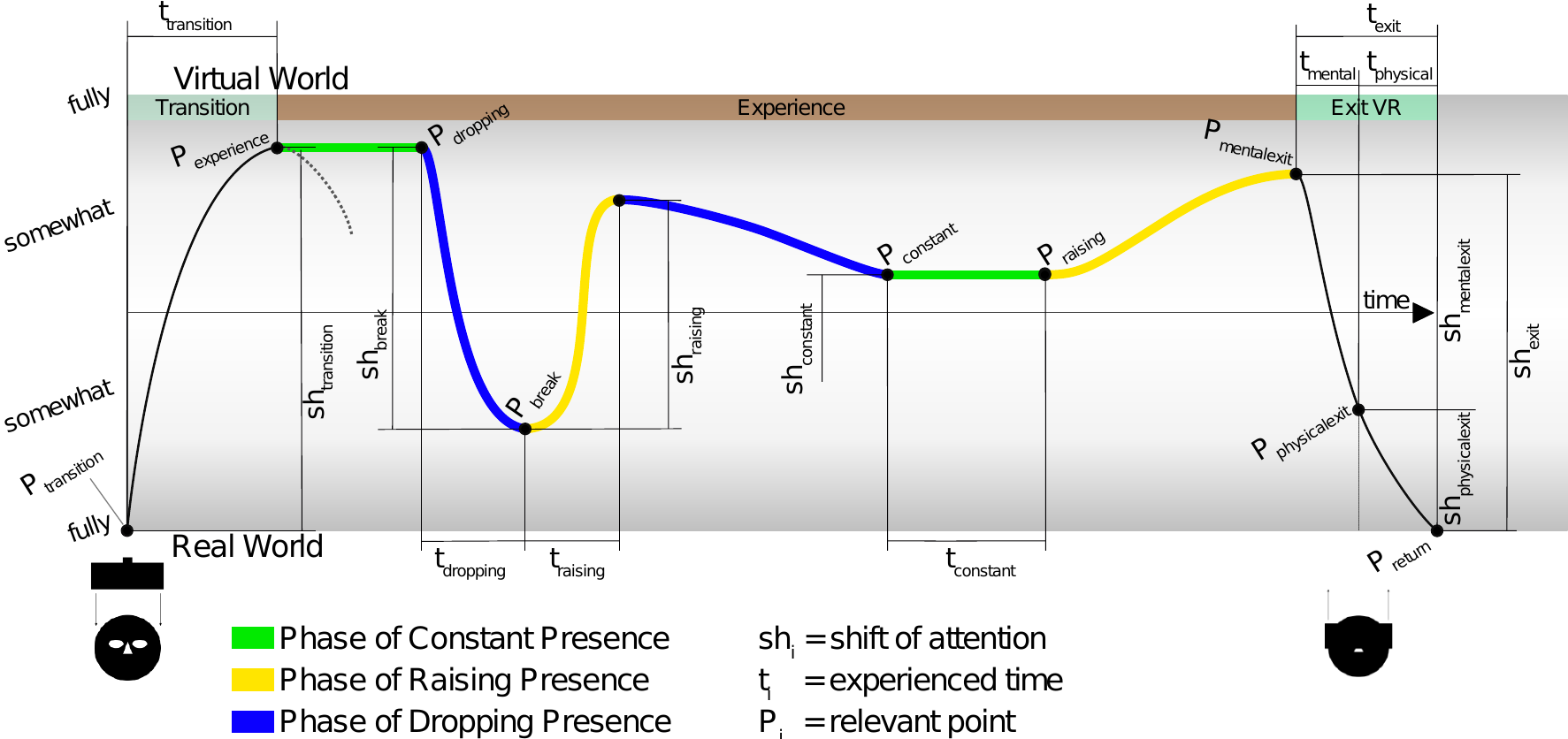}
    \caption {The drawing shows the phases experienced during a VR experience. For reasons of readability, the description shows only one example for each point and parameter.Capital letters indicate Points which define the course of the experience. Lower case letters encode the resulting parameters.}
    \label{fig:drawingarea}
\end{figure*}

We propose to provide a scale as shown in Figure \ref{fig:drawingtemplate}, with the time axis indicating the middle between being fully present in the virtual and being fully present in the real world.
By introducing a middle line, we provide a balanced diagram, which is equidistant to the extremes of being fully present in the real or the virtual world.
Further, the middle line represents a common description by users of the transition from the real to the virtual world or going over to another world.
The y-axis we chose to be 40mm in the positive and -40cm in the negative direction.
The time-axis was chosen to be 200mm between the point P\textsubscript{transition} and the P\textsubscript{physical exit} (Figure \ref{fig:drawingarea}).
The end of the experience is indicated by a dashed line with a symbol below it, indicating the time of HMD removal. 
In Section \ref{sec:model} we define and discuss the defining points in more detail.

As in the \emph{ex post facto} confrontation suggested by Garau~\cite{Garau.2008}, the remembrance of variations in presence can be supported by adding event ticks to the timeline. 
Ideally, these event ticks have the same order and relative distribution on the timeline as they appeared during the VR experience.
The event ticks might influence how the users draw their lines as they feel the need to "fit" the lines between the ticks.
Hence, depending on the evaluation objective, the examiners must decide whether a free drawing or the support of remembrance through the presentation of event ticks is more important to them.
In an exploratory VR experience, it is more difficult to provide markings of events along the timeline in advance.
In this case the supervising person can manually note done some significant events during the experience and mark it on the timeline.
Another approach would be to take screenshots throughout the experience and present them to the participant afterwards.
In the later case the ticks and according screenshots are ideally presented on a digital drawing display as it can adapt automatically to the previous experience.

The template can be used to collect general feedback for the experience or get a detailed view on single events.
In this work, we use the method as a usability tool to detect breaks in the presence experience.

First the evaluate marks descriptive points P\textsubscript{i} (Figure \ref{fig:drawingarea}, black points).
Then all points are measured and the value is stored in a data base, e.g., a table.
The position along the time-axis starts with 0mm at the line indicating putting on the HMD.
The presence state is measured from the middle line with positive values towards the virtual world and negative value towards the real world, in our case 40 and -40mm.

As drawing areas might vary in their dimensions, we normalize the values collected in the previous step.
Normalization is done by dividing the measured position for time by the distance between the lines indicating \emph{taking on the HMD} and \emph{taking of the HMD}, in our case 200mm.
The presence measure is divided by the distance between the middle line and the maximum positive distance, in our case 40mm.
From these values, the parameters t\textsubscript{i} and sh\textsubscript{i} (Figure \ref{fig:drawingarea}) are calculated by subtracting the presence values of the according P\textsubscript{break} from P\textsubscript{dropping}, which leads to a negative value.

For statistical analysis, we argue to treat the drawing as an ordinal scale.
In particular, the presence values is a continuous scale.
Therefore it is not guaranteed that the participants will use all values of the scale equidistant.
For instance a change from 0.1 to 0.5 might express less impact on the users experience then the change from 0.5 to 0.6.
Further one participant might experience a value of 0.6 as very high presence and another participant as low presence.
The user study further down shades some light on these effects in the usage of the drawing template.

\section{The Descriptive Model}
\label{sec:model}

Theoretical and practical insights influence our proposal of a descriptive model about temporal variations of a presence experience.

From the literature we derive the \textbf{three main phases} of (1) transitioning into VR~\cite{Garau.2008,OberdorferFL18,Slater:2000,sproll13,4811024,Valkov:2017:SIB:3131277.3132183}, (2) the experience itself \cite{Garau.2008,Slater.2009,Slater:2000} and (3) the exit from VR~\cite{Knibbe:2018:DCE:3173574.3174057}.
To visualize our descriptive model, we use the template introduced in the previous section (Figure~\ref{fig:drawingtemplate}).

In the following, we define starting and ending points P\textsubscript{i} of the phases, phases itself and events.
Points P\textsubscript{i} have absolute values in the coordinate system.
Parameters t\textsubscript{i} (time) and sh\textsubscript{i} are relative values between the points P\textsubscript{i}.
The parameters sh\textsubscript{i} are called \emph{shift of attention}, as they describe the cognitive shift of attention~\cite{Liebold2017}.

\subsection{Phases of a VR Experience}
The phases of every VR experience are the \emph{transition into}, the \emph{experience} itself and the \emph{exit out of VR} (Figure \ref{fig:drawingarea}).

\emph{Transition into VR} -- 
The transition into the virtual world can be divided into a physical and a virtual transition~\cite{Sungchul18,sproll13}.
During the physical transition, the HMD user habituates with the hardware, by taking on the HMD~\cite{Sungchul18,mai2018perdis,sproll13}.
The virtual transition follows the habituation, when the user can perceive a virtual picture \cite{Sungchul18, sproll13}.
The starting point of the transition could, therefore, be defined as the contact to the VR hardware or the first moment the user can perceive a virtual picture.
We argue that an HMD based experience is visually dominated (e.g.,~\cite{Kohli:2013:RT:2519692}) and one of the critical elements in a VR experience is to maintain a stable sensomotory loop~\cite{Slater.2009}, mainly based on the users' visual perception.
Also, Slater describes that the transition is not accompanied by a distinct perception of the user~\cite{Slater.2009} that could be used as a defined point in the course of an experience. 
Taking on the HMD is a conscious action that marks a distinct point in time, clearly identifiable for a supervising person and a participant.
Habituating with the hardware will influence the transition phase~\cite{Sungchul18,mai2018perdis,sproll13}, but does provoke an reaction in the user's state of presence.
Hence, we argue that the actual transition starts as soon as the HMD user puts on the HMD and is able to see a picture of the virtual world.
Point P\textsubscript{transition} marks the start of transitioning (Figure \ref{fig:drawingarea}).
The end of the transitional phase is defined by any point P\textsubscript{i}, indicating a change in the course of presence.
In our exemplary Figure~\ref{fig:drawingarea} a constant phase (green) ends the transition.

\emph{The VR Experience} - 
The experience phase is the summary of all temporal variations and phases of constant presence while acting in VR, described in the following subsection in more detail.
The first event P\textsubscript{i} defines the start of the VR experience, P\textsubscript{experience}, after the transition phase.
The end of the VR experience is the beginning of the last phase of dropping presence, indicated by the point P\textsubscript{mentalexit}.
We could not define a presence parameter for the experience phase.
The experienced time is described by t\textsubscript{experience} (Figure \ref{fig:drawingarea}, brown phase).

\emph{Exiting VR} -- 
The transition out of VR can be considered as the final drop in presence~\cite{Knibbe:2018:DCE:3173574.3174057}.
The difference is that the VR experience ends after the VR exit phase.
We assume that the \emph{VR exit} phase follows a similar process as the transition phase but in reverse order. 
We base the assumption of the reverse order on Liebold~\cite{Liebold2017} description of the transition out of VR as a conscious act, which is in line with Slater description of a BIP \cite{Slater:2000}.
The transition out of VR is a sudden realization of the user that they just had focused on a virtual environment~\cite{Liebold2017}. 
Knibbe~\cite{Knibbe:2018:DCE:3173574.3174057} collected user reports about their experience of exiting VR and found the phases, first, a mental transition -- getting aware that the VR experience ends -- and then a physical transition -- taking off the headset --.
However, they also found reports by users that describe the moment of taking off the headset as the end of the experience.
Therefore we assume both types of drawings to be valid.
As we argue that the exit might follow a reverse order to the transition into, we add the moment of taking off the headset and being able to perceive the real world visually as a defined point to the model.
The start of the \emph{VR exit} phase we define as the beginning of the last \emph{phase of dropping presence} which then is called P\textsubscript{mentalexit} (Figure \ref{fig:drawingarea}).
The end of the \emph{mentalexit} is defined by the end of the user's visual immersion.
The start of the P\textsubscript{physicalexit} is always situated on the crossing of the drawn line and the auxiliary line.
At P\textsubscript{physicalexit} the user might not be fully present in the real world yet.
By this, we take into account reports of possible aftereffects like the need for reorientation~\cite{Knibbe:2018:DCE:3173574.3174057}.
\emph{VR exit} phase ends at the point P\textsubscript{return}, describing the moment when participants report being fully present in the real world, again.

\subsection{Events in the Progress of a VR Experience}
The variation of experienced presence in a VR experience is a series of phases with constant and changing presence.

\emph{Constant Phase} --
Constant phases we define as phases without a change in the reported presence level.
In other words, the line representing the presence state over time has a slope close to zero.
We define the start of the phase of constant presence as the first point on the curve that does not show a change in presence towards the one before it.
The end of a phase of constant presence we define as the last point on the curve before the slope changes from 0.



\emph{Raising Phase} --
The start of a phase of raising presence we define as the first point on the curve that does not show a slope. The end of the \emph{phase of constant presence} can be any event P\textsubscript{i}.

\emph{Dropping Phase} --
The start of a phase of dropping presence we define as the first point on the curve that shows a negative slope. The end of the phase of constant presence can be any event P\textsubscript{i}.

\subsection{Parameters}
Parameters are relative measures between two relevant points P\textsubscript{i} and are represented in Figure \ref{fig:drawingarea} by lower case letters.
The proposed concept includes known parameters, reveals new parameters and offers the opportunity to derive new measure for the presence experience, e.g., ratios. 

People need different lengths of time to feel present in the virtual environment, which is represented by t\textsubscript{transition}.
Breaks in Presence (BIPs) \cite{Slater:2000} are represented by the Point P\textsubscript{break} and the parameter sh\textsubscript{break}.
Chung~\cite{Chung:2009:MTV:1670252.1670287} defined a parameter similar to sh\textsubscript{break} as the intensity of an attention shift.

The parameter t\textsubscript{dropping} helps to differentiate between sudden BIPs, e.g., a loud noise from the real world, or longer lasting effects, e.g., a tedious experience that makes the users attention drift slowly back to the real world over time~\cite{Garau.2008,Liebold2017} (compare exemplary Figure \ref{fig:drawingarea} first and second phase of dropping presence).
Further, the relation between sh\textsubscript{break} and a following t\textsubscript{raising} gives insights into the intensity of the break as recovery time was shown to be slightly correlated with the intensity of the break \cite{chung12}.
We expect, according to Slater \cite{Slater:2000}, that t\textsubscript{dropping} in general will be shorter then t\textsubscript{raising}.
t\textsubscript{transition} is expected to be longer than t\textsubscript{exit}.

For the phase of exiting VR we define the three temporal parameters, total time t\textsubscript{exit}, mental transition t\textsubscript{mental} and t\textsubscript{physical}, as argued in the subsection \emph{Exiting VR} above.

\section{Exploration of the Method}
\label{lab:exploration}
In the previous sections we introduced a method to collect and store user reports about temporal fluctuations in the presence experience during an HMD session.
As we propose, the method can be used as a method to detect breaks in presence (BIP) in an experience, a tool evaluate temporal fluctuations in more detail or a tool to compare experiences against each other.
In this work, we focus on showcasing the feasibility of the method and the usage as a BIP detector.
Further, this exploratory study gives insights in the usage of the drawing template and the possibilities to store and evaluate data.
We conducted a controlled lab experiment to gain insights in the application of the drawing method and the descriptive model. 
The experience was designed in such a way that, using classical methods, it no longer showed any breaks in presence. 
We then introduced different kinds of BIP with varying strength.
The goal of the exploration is (1) to gain insights into the repeatably of the drawing method between users and (2) the quality of detecting and evaluating BIPs.

\subsection{Measures}
\label{lab:measures}
To gather insights in the (1) repeatably we argue that the drawings need to follow a certain progress.
As the drawings will have strong interpersonal variations we will look for particular events that are known from related work (see Section~\ref{sec:model} for more detail). 
Hence we will analyze the drawings on key characteristics, we defined in the descriptive model (Figure~\ref{fig:drawingarea}).
The characteristics we expect a user to draw are:
\begin{itemize}
\setlength{\itemsep}{0pt}
\setlength{\parskip}{0pt}
\setlength{\parsep}{0pt}
\item[a:] Lines will start at the point P\textsubscript{transition}.
\item[b:] Point P\textsubscript{return} exists.
\item[c:] Break in presence are dropping lines (sh\textsubscript{break}).
\item[d:] The experience phase will cover most of the time axis.
\item[e:] An attention shift towards the virtual world takes a longer time span then towards the real world.
\end{itemize}

If the majority of participants follows this structure, our method is feasible of analyzing VR experiences.

To gain insights on the possibilities of the method as a BIP detector (2) we will count the detected breaks. Furthermore, we analyze possible order effects, e.g. how the early occurrence of a BIP affects the possibility to get reported. 
Finally, we gain insights in the possibility to reflect the strength of a BIP.
Additionally, we collected qualitative feedback to gain better insights into the understanding and usage of the drawing method by the participants.

\subsection{Task}
\begin{figure}
    \centering
    \includegraphics[width=\columnwidth]{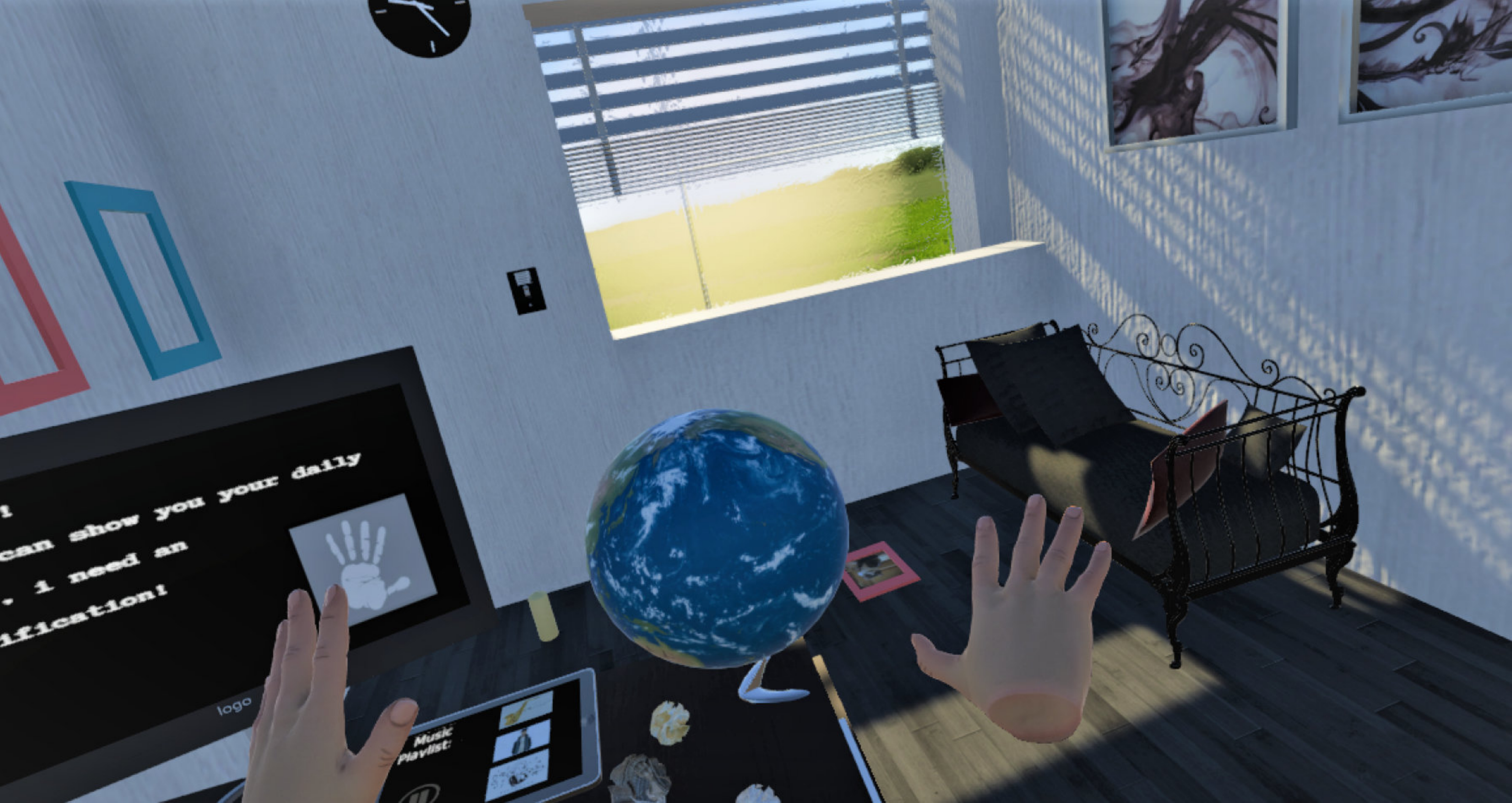}
    \caption {Perspective of the user when entering the room.}
    \label{fig:egoview}
\end{figure}

The proposed drawing method is thought to detect BIPs.
We designed a linear game consisting of six subtasks to be solved by the user that did not include any BIPs.
The dimensions of the virtual room were $3.5 x 2.5 meters$ and fitted into the lab space with $4.0 x 3.0 meters$ (Figure \ref{fig:egoview}).
In three iterative design steps with a total of 14 participants and four experts, we removed all unintended BIPs. 
We used the virtual experience test~\cite{Chertoff:2010:VET:2195920.2196260} and the igroup presence questionnaire (IPQ)~\cite{Schubert2003} during the design process.
In addition, we conducted semi-structured interviews asking the participants for the positive and negative aspects of the experience as well as reminders of the real world and flaws in the design of the virtual world. 
In a final evaluation with eight novice participants the experience was rated to have a high presence on the IPQ ($G1: 4.6$; $SP:5.1$; $INV: 4.8$; $REAL: 3.1$) and no events influencing the experience were reported.

During the experience, the participants fulfilled six tasks.
The first was a paper-tossing, in which they threw paper balls in a garbage bin.
The second was to find two pictures of cats in the room and hang it on the corresponding spot on the wall.
In the third participants had to find and tidy up pens.
In the fourth, the goal was to find books and put them in the right order into a bookshelf to complete the word reality.
In the fifth, they had to solve a riddle which solution was to use a switch on the wall.
The sixth task was also a riddle which solution was to turn on the music and make it play Beethoven.

\subsection{Stimuli -- Breaks in Presence}
\label{sec:BIPS}
As stimuli we designed BIPs.
We varied the intensity, which is a combination of violations of the users mental model about the VR and the strength of the distracting stimulus, which is described in detail by Liebold~\cite{Liebold2017}.
The duration was chosen to be between 2 and 4 seconds, according to related work~\cite{Chung:2009:MTV:1670252.1670287,Garau.2008}.
The breaks targeted spatial presence, realness, and involvement as defined by Schubert~\cite{Schubert2003} and affected all senses.

We introduced five BIPs that are ordered in the following list from the strongest to the weakest:
\begin{enumerate}
    \item BIP Cable Malfunction (BIPCM)
    A four second long \emph{blue screen} error is shown in the HMD and the examiner asks the participants verbally not to move, followed by touching and moving the cable attached to the HMD to "solve" the issue.
    
    \item BIP Whitescreen (BIPWHITESCREEN)
    A four-second long white screen with a beeping audio track was shown similar to Garau \cite{Garau.2008}.
    
    \item BIP Teleport (BIPTELEPORT)
    The users are teleported for four seconds to an empty virtual room, having the same dimension as the main room, but showing pictures of the physical room surrounding them on the wall.
    
    \item BIP Failed Interaction (BIPFAIL)
    Users' had to put two framed pictures of a cat on the wall by holding it next to a marking on the wall and releasing the grasp. The second picture was made to stick after the seventh attempt to put the picture on the wall. Until the seventh attempt, the picture will fall, or the user might catch it in the air.
    
    \item BIP Unrelated Controller Vibrations (BIPVIBRATION)
    The right controller vibrated for four seconds when it was not used for any purpose.
\end{enumerate}

The breaks were triggered between 5 and 10 seconds after a task was solved.
The BIPs appearance was randomized in three groups.
The concept was to have a weak and strong break at the beginning and the end in the sequence of BIPs, in one condition each.
Further the strong breaks were clearly separated once and one time close together.
The resulting randomization orders are: A (2-4-5-3-1), B (5-4-2-1-3) and C (1-4-3-5-2).

\subsection{Apparatus}
The setup consisted of an HTC Vive HMD with according controllers\footnote{\url{https://de.wikipedia.org/wiki/HTC_Vive}}
, the TP-Cast system wireless system for the HTC Vive \footnote{\url{https://www.tpcastvr.com}}. 
The participant used Bose QC25 Noise Cancelling Headsets.
The virtual experience was generated and run at 90 frames per second using Unity 5.6 on a desktop PC with NVIDIA GTX 1080 graphics card and an Intel I7 processor. 
The presented virtual scene was optimized to evoke presence according to the five parameters introduced by Schubert \cite{schubert_experience_2001}.
\emph{Dramatic Involvement} was realized by giving the user the overall task to clean up the office, as the boss will arrive soon.
\emph{Quality of Immersion} was addressed by optimizing the visual appearance of the VR scene regarding 3D models, texturing and lighting. In addition, spatialized sound was used with three ambient sound sources and twenty audio sources.
The controllers vibrated when an object was touched.
To support \emph{Exploration of VE}, all objects, but heavy furniture was movable or reactive to actions (e.g., spinning a globe on the desk, Figure \ref{fig:egoview}) and behave physically to support \emph{Predictability}.
\emph{Interface Awareness} was reduced by enabling basic interactions with just one button all the time -- the users had virtual hands that supported a grasping animation --, and a wireless HMD.

\subsection{Participants}
30 participants (11 females, 19 males, mean age= 24) that did not take part in the iterative design process were invited by mailing lists, paper displays, social media, and personal social network. 
The participants had backgrounds from diverse fields. Twelve participants had never used a VR headset before, the rest reported to have used HMDs less than one time a year.
None of the participants had used the drawing template before. 
Participants were rewarded study credits or 10 euro according to their choice.

\subsection{Procedure}

Participants were welcomed, and the purpose of the study was explained as being a user experience study about interaction in VR. 
The whole study took about 60 minutes, whereby approximately 15 minutes were spent in VR.
The participants were informed about anonymous storage and evaluation of their movement data given by the HTC Vive tracking system, recording of screen capture and sound in the room and asked to fill out a consent form. 
The usage of the HMD, noise canceling headset and controllers was explained to and tested by the participant in the standard grey Steam VR environment.
As soon as the participants felt comfortable in using the system, they took off the HMD again, were guided to a starting position marked on the floor.
The participants were instructed that the examiner will not help them during the experience.
They were informed that the session will end after the last task and a prompt will show up, indicating the end of the experience.
No information about possible BIPs was given.
They started the experience on their own by taking on the HMD.
The first thing the participants saw, was a transitional room.
The transitional room had the same size and layout as the final room, but only a few objects were presented on a desk in front of the user. 
The objects were used to again train the interaction in VR.
The participants were guided by instructions, which were presented textually on a monitor on the desk, during the whole experience. 
After completing the tutorial, users were teleported into the main room. By pressing a virtual button they activated the teleport themselves.

After the experience, the participants conducted the drawing task as described in Section \ref{sec:drawingmethod}.
When finished, the examiner conducted an interview about the graphs to record details about the general course of presence and to assign the appropriate drawn BIPs to the actual ones.
Finally, the participants were asked to fill out the IPQ \cite{Schubert2003} and demographic information about age, sex and previous experience with HMDs.

\subsection{Results}

\begin{figure}
    \centering
    \includegraphics[width=\columnwidth]{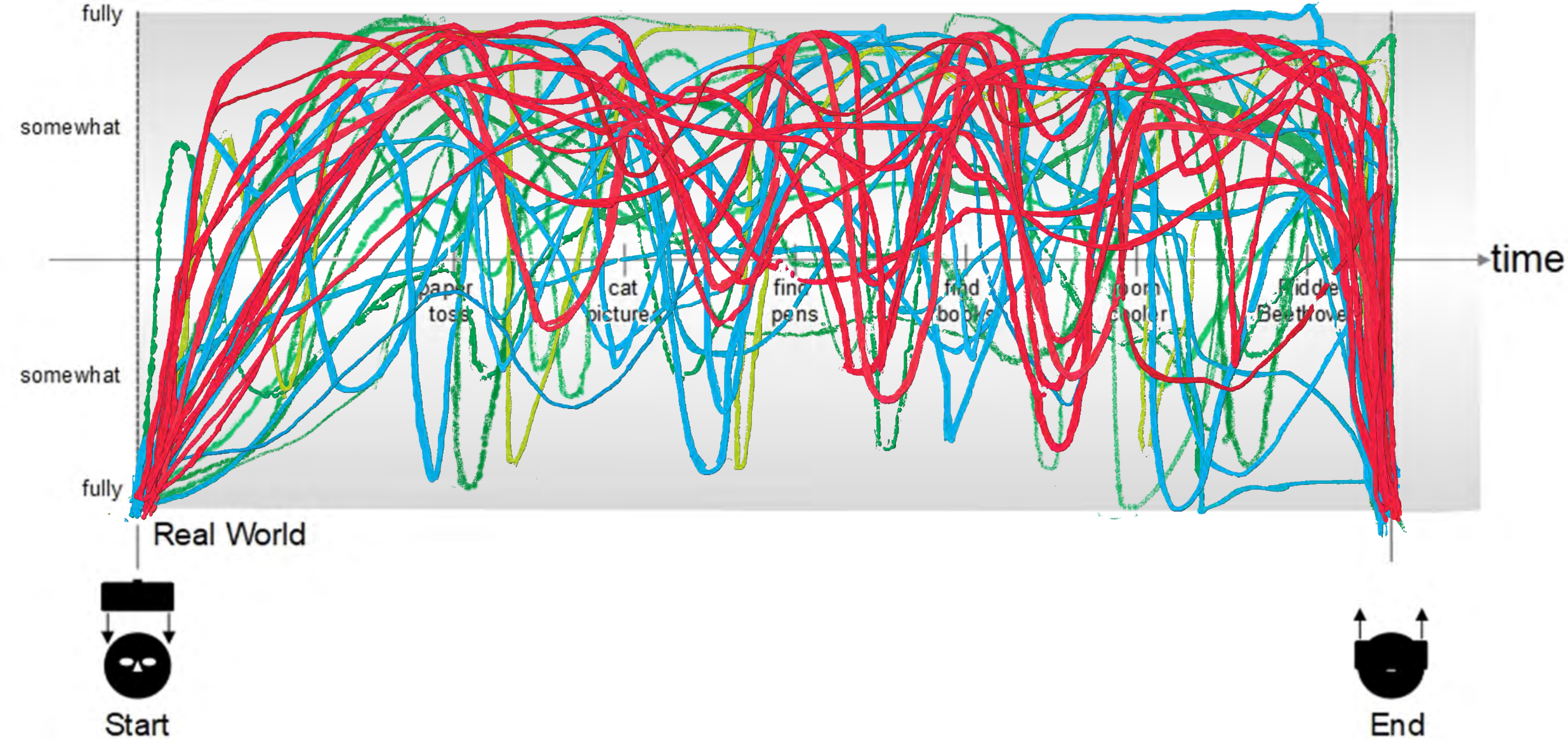}
    \caption {All drawings collected in the exploratory study. Annotations removed for better readability. The figure illustrates the need for our descriptive model based on numerical values and logical values that enable a comparison independent of the drawings. Colours represent the randomization groups ($green=A, red=B, blue=C$)}
    \label{fig:allgraphs}
\end{figure}

The results section is thought to answer our research questions.
Additionally we showcase how to use and analyze the data by the help of the method and the descriptive model we proposed.
To get a better understanding for the method we will present the data in different levels of detail.
We propose analyses that are useful to answer our research question, but future researchers are open to work with their data in their own way. 

\begin{figure}
    \centering
    \includegraphics[width=\columnwidth]{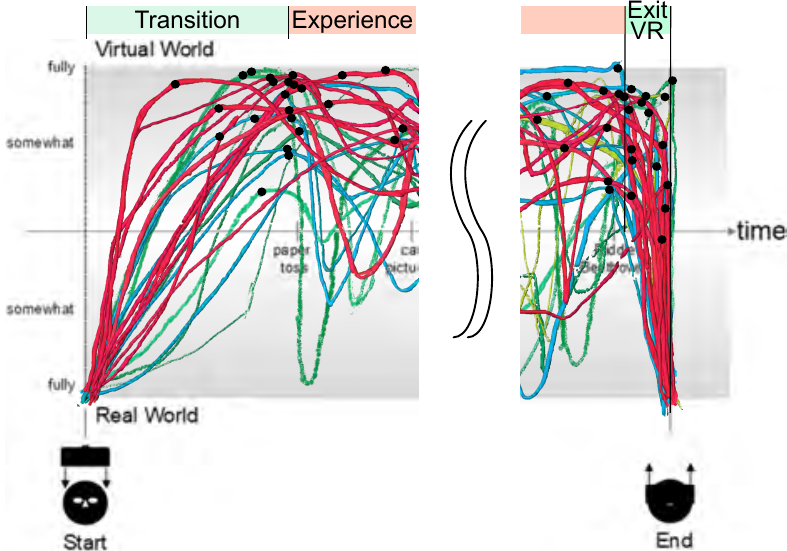}
    \caption {Transition into VR is reported to be longer than the transition out of VR, which is in line with the related work. Black dots show the transition points P\textsubscript{experience} (left) and P\textsubscript{mentalexit} (right). (Randomization groups: $green=A, red=B, blue=C$)}
    \label{fig:Transitionalphases}
\end{figure}

Figure~\ref{fig:allgraphs} shows the drawing collected during the exploratory study.
It showcases the diversity in user drawings.
Since all the randomization groups are shown in the figure, it represents the possible result of a real application where the occurrence of BIPs is unpredictable.
To obtain data that can be analyzed, we use our proposed method and the descriptive numerical model.
Exemplary Figure~\ref{fig:Transitionalphases} shows the marking of the points indicating the end of transitioning into and the begin of exiting VR. 
The values of these points are stored in the database together with the annotations describing them.
We were able to store all points with the help of the proposed descriptive model and stored them the database.
One P\textsubscript{return} could not be described by our model as the end of the line did not show a position that is defined as being fully present in the real world and therefore was excluded.

Tabel~\ref{tab:frequency} presents the outcome of a statistical evaluation of the drawings based on the database for the single randomization groups. 
It shows the evaluation of points (P\textsubscript{break}) as well es derived parameters (sh\textsubscript{break}).
Figure~\ref{fig:BIPsoverall} is en example for the visualization of the numerical data. 
It shows Box-Plots for the introduced BIPs, combining the results of all randomization groups.

\begin{table*}
\begin{center}
\begin{tabular}{lcccccccccccc}
                                                                                       & \multicolumn{12}{c}{Randomization}                                                                                                                                                                                                                                                                                                                                                                                                                                                                                                                                                                                                                                                                                                                                                                        \\ \cline{2-13} 
\multicolumn{1}{l|}{}                                                                  & \multicolumn{4}{c|}{A}                                                                                                                                                                                                                                            & \multicolumn{4}{c|}{B}                                                                                                                                                                                                                                            & \multicolumn{4}{c|}{C}                                                                                                                                                                                                                                            \\ \hline
\multicolumn{1}{|l|}{\begin{tabular}[c]{@{}l@{}}Breaks in \\ \\ Presence\end{tabular}} & \multicolumn{1}{c|}{POS} & \multicolumn{1}{c|}{\begin{tabular}[c]{@{}c@{}}Detect.\\ (\%)\end{tabular}} & \multicolumn{1}{c|}{\begin{tabular}[c]{@{}c@{}}$sh_{break}$\\\end{tabular}} & \multicolumn{1}{c|}{\begin{tabular}[c]{@{}c@{}}$P_{break}$\\ \end{tabular}} & \multicolumn{1}{c|}{POS} & \multicolumn{1}{c|}{\begin{tabular}[c]{@{}c@{}}Detect.\\ (\%)\end{tabular}} & \multicolumn{1}{c|}{\begin{tabular}[c]{@{}c@{}}$sh_{break}$\\ \end{tabular}} & \multicolumn{1}{c|}{\begin{tabular}[c]{@{}c@{}}$P_{Break}$\\ \end{tabular}} & \multicolumn{1}{c|}{POS} & \multicolumn{1}{c|}{\begin{tabular}[c]{@{}c@{}}Detect.\\ (\%)\end{tabular}} & \multicolumn{1}{c|}{\begin{tabular}[c]{@{}c@{}}$sh_{break}$\\ \end{tabular}} & \multicolumn{1}{c|}{\begin{tabular}[c]{@{}c@{}}$P_{Break}$\\ \end{tabular}} \\ \hline
\multicolumn{1}{|l|}{\begin{tabular}[c]{@{}l@{}}Cable\\ Malfunction\end{tabular}}      & \multicolumn{1}{c|}{5}   & \multicolumn{1}{c|}{100}                                                    & \multicolumn{1}{c|}{\begin{tabular}[c]{@{}c@{}}-0.33\end{tabular}}     & \multicolumn{1}{c|}{\begin{tabular}[c]{@{}c@{}}-0.72\end{tabular}}   & \multicolumn{1}{c|}{4}   & \multicolumn{1}{c|}{90}                                                     & \multicolumn{1}{c|}{\begin{tabular}[c]{@{}c@{}}-0.28\end{tabular}}     & \multicolumn{1}{c|}{\begin{tabular}[c]{@{}c@{}}-0.33\end{tabular}}   & \multicolumn{1}{c|}{1}   & \multicolumn{1}{c|}{70}                                                     & \multicolumn{1}{c|}{\begin{tabular}[c]{@{}c@{}}-0.28\end{tabular}}     & \multicolumn{1}{c|}{\begin{tabular}[c]{@{}c@{}}-0.34\end{tabular}}   \\ \hline
\multicolumn{1}{|l|}{\begin{tabular}[c]{@{}l@{}}White \\ Screen\end{tabular}}          & \multicolumn{1}{c|}{1}   & \multicolumn{1}{c|}{60}                                                     & \multicolumn{1}{c|}{\begin{tabular}[c]{@{}c@{}}-0.38\end{tabular}}     & \multicolumn{1}{c|}{\begin{tabular}[c]{@{}c@{}}-0.4\end{tabular}}   & \multicolumn{1}{c|}{3}   & \multicolumn{1}{c|}{70}                                                     & \multicolumn{1}{c|}{\begin{tabular}[c]{@{}c@{}}-0.25\end{tabular}}     & \multicolumn{1}{c|}{\begin{tabular}[c]{@{}c@{}}-0.03\end{tabular}}   & \multicolumn{1}{c|}{5}   & \multicolumn{1}{c|}{70}                                                     & \multicolumn{1}{c|}{\begin{tabular}[c]{@{}c@{}}-0.28\end{tabular}}     & \multicolumn{1}{c|}{\begin{tabular}[c]{@{}c@{}}-0.29\end{tabular}}   \\ \hline
\multicolumn{1}{|l|}{Teleport}                                                         & \multicolumn{1}{c|}{4}   & \multicolumn{1}{c|}{30}                                                     & \multicolumn{1}{c|}{\begin{tabular}[c]{@{}c@{}}-0.45\end{tabular}}      & \multicolumn{1}{c|}{\begin{tabular}[c]{@{}c@{}}-0.15\end{tabular}}   & \multicolumn{1}{c|}{5}   & \multicolumn{1}{c|}{50}                                                     & \multicolumn{1}{c|}{\begin{tabular}[c]{@{}c@{}}-0.18\end{tabular}}     & \multicolumn{1}{c|}{\begin{tabular}[c]{@{}c@{}}0.08\end{tabular}}    & \multicolumn{1}{c|}{3}   & \multicolumn{1}{c|}{40}                                                     & \multicolumn{1}{c|}{\begin{tabular}[c]{@{}c@{}}-0.3\end{tabular}}     & \multicolumn{1}{c|}{\begin{tabular}[c]{@{}c@{}}-0.11\end{tabular}}   \\ \hline
\multicolumn{1}{|l|}{\begin{tabular}[c]{@{}l@{}}Failed\\ Interaction\end{tabular}}     & \multicolumn{1}{c|}{2}   & \multicolumn{1}{c|}{70}                                                     & \multicolumn{1}{c|}{\begin{tabular}[c]{@{}c@{}}-0.2\end{tabular}}     & \multicolumn{1}{c|}{\begin{tabular}[c]{@{}c@{}}-0.23\end{tabular}}   & \multicolumn{1}{c|}{2}   & \multicolumn{1}{c|}{90}                                                     & \multicolumn{1}{c|}{\begin{tabular}[c]{@{}c@{}}-0.15\end{tabular}}      & \multicolumn{1}{c|}{\begin{tabular}[c]{@{}c@{}}0.23\end{tabular}}    & \multicolumn{1}{c|}{2}   & \multicolumn{1}{c|}{50}                                                     & \multicolumn{1}{c|}{\begin{tabular}[c]{@{}c@{}}-0.34\end{tabular}}     & \multicolumn{1}{c|}{\begin{tabular}[c]{@{}c@{}}-0.2\end{tabular}}   \\ \hline
\multicolumn{1}{|l|}{Vibration}                                                        & \multicolumn{1}{c|}{3}   & \multicolumn{1}{c|}{20}                                                     & \multicolumn{1}{c|}{\begin{tabular}[c]{@{}c@{}}-0.1\end{tabular}}      & \multicolumn{1}{c|}{\begin{tabular}[c]{@{}c@{}}0.19\end{tabular}}    & \multicolumn{1}{c|}{1}   & \multicolumn{1}{c|}{0}                                                     & \multicolumn{1}{c|}{\begin{tabular}[c]{@{}c@{}}-\end{tabular}}      & \multicolumn{1}{c|}{\begin{tabular}[c]{@{}c@{}}-\end{tabular}}    & \multicolumn{1}{c|}{4}   & \multicolumn{1}{c|}{20}                                                      & \multicolumn{1}{c|}{-0.22}                                                      & \multicolumn{1}{c|}{0.09}                                                     \\ \hline
\end{tabular}

\end{center}
\caption{The results for detection rate and reported intensity of the BIP in the three randomizations. The strongest -- Cable Malfunction -- is in the top row, the weakest -- controller vibration -- in the bottom row. POS describes the appearance in each visualization group, whereas 1 is first place and 5 is the last place. The values sh\textsubscript{break} P\textsubscript{break} (see Section~\ref{sec:model} and Figure~\ref{fig:drawingarea} for more details) report the averages.}
\label{tab:frequency}
\end{table*}

\subsubsection{The Transitional Phases}
Figure \ref{fig:Transitionalphases} shows the drawn transition and exit phases, without the experience in between.
Six participants experienced a disturbance during transition, not caused by the BIPs we introduced intentionally. 
We exclude them in Figure~\ref{fig:Transitionalphases} (left picture).
The unintended disturbances are caused by the HMD not fitting and readjustments by the participants.
One very shallow transition drawing was caused by boredom during the instructional part, leading to a low feeling of involvement.
The last one did not give a justification.
Exiting VR was influenced for one participant of Randomization C as the \emph{BIPWHITESCREEN} happened as the last break before the end of the experience.  
It was removed in Figure~\ref{fig:Transitionalphases} (right picture) as no exit phase was drawn according to our definition.
The excluded cases are still visible in the overview of drawings (Figure~\ref{fig:drawingarea}).

Further, Figure~\ref{fig:Transitionalphases} is an example for the application of the analysis (Figure~\ref{fig:teaser}).
As we are able to store the data in a database, we can calculate parameters, for instance on the experience of time.
In average participants used $21\%$ ($SD=10\%$) of the timeline to report the transition into VR (t\textsubscript{transition}).
Further they used $8\%$ ($SD=5\%$) to express the transition out of VR (t\textsubscript{exit}).
As a result the \emph{experience} phase took 71\% of the available space.
P\textsubscript{transition} was always situated at the defined position, as it was an instruction.
Participants finished their drawings in average at P\textsubscript{return} $99.6\%$ ($SD=1\%$) of the timeline, and at $-93\%$ ($SD=30$) of the negative presence axis.

In no cases, participants reported separating their exit experience in the \emph{mental} and \emph{physical} exit.
25 of the drawings show an drop which accelerates continuously over time and shows a uniformly accelerated slope of the curve between the points P\textsubscript{mentalexit} and P\textsubscript{return}.
Five participants drew their experience as a sudden breakpoint at which P\textsubscript{mentalexit} is at the same value on the timeline as P\textsubscript{return}.

\subsubsection{Breaks in Presence}

\begin{figure}
    \centering
    \includegraphics[width=\columnwidth]{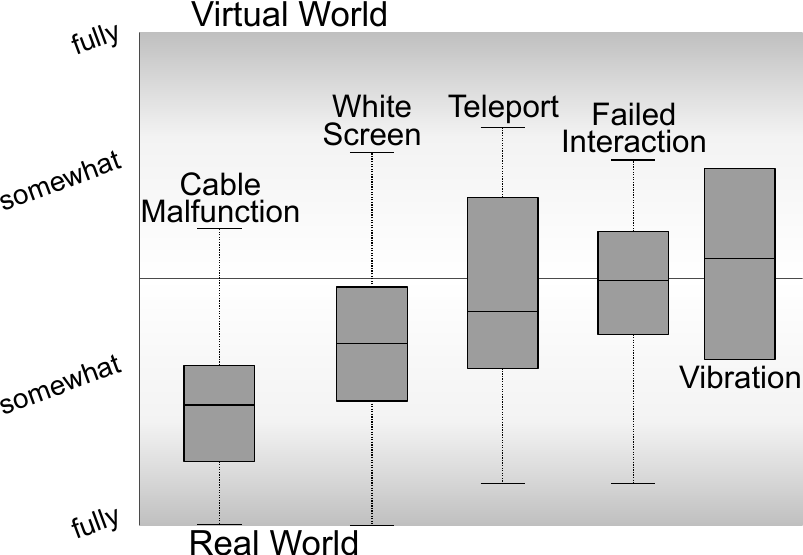}
    \caption {Absolute BIP intensity average for all randomization. The strength of the breaks is reported according to our assumptions, which indicates that drawings give an estimate for the intensity of a break.}
    \label{fig:BIPsoverall}
\end{figure}

\emph{Detection Frequency} -- The detection frequency is the percentage of participants reporting a BIP.
We report the detection frequencies in Table~\ref{tab:frequency}.
Strong BIPs -- \emph{Cable Malfunction}, \emph{White Screen} -- are more likely to be reported than weak BIPs, no matter if they appear early or late during the experience.
\emph{BIPFAIL} stands out from this with a high detection rate in two randomization groups.
Although, we defined it as a weak BIP and the disturbance resulting from it was perceived by the participants as such during the experiment (Figure~\ref{fig:BIPsoverall}), the detection rate was similar to the stronger BIPs.
An individual participant was able to identify 2.8 BIPs ($SD=0.9$) on average and across all randomization groups.

\emph{Order effects} --
The order might show an effect in our study, but the results are not clear as the number of detections was low for some BIPs, which prevents further statistics.
The two strongest breaks are 10\% and 40\% more likely to be reported when in the last position of the BIPs compared to being the first.
However the weaker breaks do not show such an effect.
The BIP \emph{failed interaction}, which is in the same position of the sequence of BIPs in all randomizations, shows a variation in the detection rate between 50\% to 90\%.
The unrelated vibration of the controller was not reported, when in the first position of the randomization -- Position 1, Randomization B --.

\emph{Intensity} -- 
Figure~\ref{fig:BIPsoverall} shows Box-Plots for the presence value of P\textsubscript{break} considering all randomization groups as one.
The intensity for P\textsubscript{break} in the single groups can be found in Table~\ref{tab:frequency} but is not used for further discussion, as the number of participants is to small to recognize the trend.
A trend in the order of the BIPs' strength is visible that fits the prediction we derived from related work in Section~\ref{sec:BIPS}.
The weaker breaks are drawn with a tendency of their median towards the \emph{virtual world}.
In particular, the strongest BIP \emph{BIPCM} is reported to be close to the real world, the weakest BIP \emph{BIPVIBRATION}.
We did not find a tendency for the parameter sh\textsubscript{break} in our study.


\emph{Time Parameters} --
At this point we want to stress that the timely order is not of particular interest for the detection of BIPs and in our study. 
However, it gives an estimate on how precise users are able to report the appearance of a BIP, if the method is used in a real world application.
We were interested in the ability to report appearance of BIPs relative to events in the experience.
We counted the correct positioning of P\textsubscript{dropping} or the following P\textsubscript{i}.
BIPs were counted as positioned correctly if one of the points was situated within 5mm (2.5\%) before and 25mm (12.5\%) after the according task tick.
We found that participants had 97 of the 118 detected breaks positioned correctly.
As a result in 18\% of the cases the relation between the event in VR and the drawing would not be clear.

We further compared the time it takes that a BIP comes into full effect, t\textsubscript{dropping}, to the time participants need to recover from it, t\textsubscript{raising}.
We calculated the average of each and then divided the result of t\textsubscript{dropping} by t\textsubscript{raising}.
The resulting ratio is 0.8, which means that a dropping phases is drawn 20\% shorter than a raising phase.

\subsection{Other Anecdotal Findings}
When drawing the curves, many users started to reflect on the drawings, stimulated by the drawn line.
Some recognized, after drawing a second BIP, that previous one was stronger.
Therefore they went back to the first and changed the curve to fit relative to the second.
We found similar behavior for ordering the appearance BIPs and their relative position to the event ticks on the timeline.

\section{Discussion}

\emph{The method enables users to reflect temporal variations} -- 
The drawings fulfill our five prerequisites (Section~\ref{lab:measures}) to report temporal variations during a presence experience.
In particular we could show that all participants started at P\textsubscript{transition} (prerequisite a) and end with P\textsubscript{return} (prerequisite b).
Only one event was drawn that did not fit into the model.
In the particular case, presence was reported as raising again after the last dropping phase (Figure~\ref{fig:drawingarea}, randomization C).
All participants understood that a BIP is an attention shift towards the real world and drew it as a dropping line (prerequisite c).
Most of the template (71\%) was used to express the temporal variation during the main \emph{experience} phase (prerequisite d).
According to the related work, we showed that participants report an attention shift towards the real world to happen faster than towards the virtual world (prerequisite e).
In particular t\textsubscript{transition} was 38\% longer than t\textsubscript{exit} and to recover from a break (t\textsubscript{raising}) took in average 20\% longer than the break (t\textsubscript{dropping}).

In summary, we were able to show that our method is superior to established methods, e.g., by using Garau's approach~\cite{Garau.2008}, when it comes to understanding temporal variations in the presence experience.
The expression was not possible in such possible before.
We want to stress at this point, that the participants had minor experience in HMD usage and were not given any information about expectations on the drawings progress.
The strong coherence between the theoretical model we provide in section~\ref{lab:measures} and the drawings, reflecting the users actual mental model, underlines the feasibility of our method.

\emph{Drawing Method is a tool to detect BIPs} -- 
The following paragraphs target research question two that we raised in Section~\ref{lab:exploration}.
The groups of ten Participants were able to detect all BIPs, but the weakest BIP when it occurred at the beginning of the experience (Table~\ref{tab:frequency}).
The two strongest breaks base on visual disturbances and are detected by at least 60\% of the participants.
We find it interesting that the failed interaction was reported by 50\% to 90\% of the users.
A failed interaction is not a technical failure disturbing the perception of the virtual world.
It is a logical failure of the experience, disturbing the plausibility illusion~\cite{Slater.2009} for the HMD user.

Furthermore, we found BIPs that we did not find during the design phase with established methods.
In particular, in the transition phases we identified 7 disturbances, with accompanying user reports on the effects on the further course of experience.
These results underline the possibilities offered by our method compared to established methods. 
Additionally, it demonstrates the advantage of being able to analyze the temporal cause of the presence experience.
The disturbance in the beginning affected the user over a long period in time, which would have been impossible to express with a questionnaire and more complex to explain in an interview.

Similar to other post-experience methods the drawing method suffers from inaccurate recall by the participants \cite{Garau.2008}.
However, in each randomization group, with 10 participants each, almost all breaks were reported (Table~\ref{tab:frequency}) by at least 20\%.
As a result, we argue that 10 participants are enough to discover all BIP in a VR experience with a maximum duration of 15 minutes.
There might be a chance to need even fewer participants, but with the risk of not finding a weak disturbance that occurs early in an experience.
We want to highlight at this point, that using 10 or less people to find errors in a system is similar to established recommendations from usability testing of other systems~\cite{Nielsen:1993:MMF:169059.169166}.
Longer experiences might demand more participants, to counteract inaccurate recall of weak breaks at the beginning of the experience.

In summary, our method is able to detect a variety of BIPs belonging to all aspects of the presence experience.
Finding BIPs that we have not consciously implemented in the experience shows the advantage of our method over established methods.

\emph{Drawings indicate the intensity of a BIP} -- 
The users reported the disturbance introduced by the BIPs (Figure \ref{fig:BIPsoverall}).
The points P\textsubscript{breaks} represent BIPs the best, which is in line we Chung's suggestion \cite{chung12}. 
We were speculating that the parameter sh\textsubscript{break} might report the relative evaluation by the participants. 
We could not find a clear trend for sh\textsubscript{break}, as it might depend on things that precede or follow.
However, future research might have a closer look on the parameter sh\textsubscript{break}, as it has the potential to analyze accumulative effects of events on the overall presence experience.

In contrast to the overall intensity for P\textsubscript{Break} (Figure~\ref{fig:BIPsoverall}), the single randomization groups do not show a clear trend for the intensity of a break (Table \ref{tab:frequency}). 
One reason might be that the lower number of participants is not enough to detect a difference.
Another be that the randomization levels out effects that are introduced by the order the breaks had in our study.

From this, we conclude that our drawing template has good potential to evaluate the intensity of a BIP.
However, the number of participants needed to evaluate the strength of a break needs to be much higher then for the detection of the breaks.
It needs to be mentioned that we decided not to remove outliers.
We decided to leave all data in the database, as the users decided to draw the line according to their believe on how to use the model.
Therefore they did nothing wrong that justifies an exclusion.
It will need further studies to gain a deeper understanding of why people chose to draw their line very differently to others to decide for outliers.


\emph{Drawing the transitional phases has potential} -- We could show that people report the experience of a transition phase in average around three times longer than the exit phase (Figure~\ref{fig:Transitionalphases}).
This is according to the assumption that people do harder to detect the change of the presence state towards the VR but the break shifting their attention towards the real world~\cite{Liebold2017,Slater:2000}. 

We could not find a separation of the mental and physical exit of VR~\cite{Knibbe:2018:DCE:3173574.3174057}.
Novice participants might be overwhelmed by the things happening in the short time during the exit.
83\% of the participant drew the exit as a continuous curve that might indicate a separation of these phases.
The rest drew a sudden break (Figure \ref{fig:Transitionalphases}).
However the continuous curves might be a result of a more natural or comfortable drawing behavior and not related to the experienced changes.
Informing the user about possible effects during the exit phase might support a more differentiated report.

As our study was not designed to evaluate the transitional phases, certain parameters might need clarification in future studies.
For example, we defined the parameter t\textsubscript{transition} to be the time between P\textsubscript{transition} and the next prominent point, as it is an ongoing process.
However, the time t\textsubscript{transition} it actually needs to develop a sufficient presence experience might be indicated by the point that crosses the middle line between real world and virtual world.

We can not answer these questions here, as they are not the main goal of hour study.
However, in conjunction with the findings discussed above about detailed user reports on their transitional phases, we see great potential to better understand the transitional phases of a VR experience when using our method.

\subsection{Limitations}
The report of a longer transition time into VR then to exit VR, is in line with the related work.
However, there might be a bias introduced by the drawing template.
In the template the ticks for the tasks -- e.g., putting a picture on the wall -- are distributed along the time axis according to their appearance relative to events in the experience.
The first task in the experience -- paper toss -- was much later in time, then it took between the last task -- Riddle Beethoven -- and the end of the experience.
Therefore, there was much more space to draw the transition into than the exit of VR, which might lead to the current result (Figure \ref{fig:Transitionalphases}).

We do not cover drawings that show a step raising at the beginning and then change to a more shallow curve when transitioning into VR.
We decided to stick to our decision and used the first point of constant presence or dropping presence for the parameters t\textsubscript{transition} and sh\textsubscript{transition}.
However, people might express that they felt present in the virtual world very fast, but became more and more confident over time, as known from other studies \cite{Mai_}.

The experience in this study took about 15 minutes.
Longer experience might amplify possible inaccurate recall of the experience.
An analysis of longer experiences should include more than 10 participants.

\section{Future Work}
The current work aimed on creating the drawing method and the proposal of a descriptive model to describe the outcome in an objective way.
Exemplary we chose the use case  to explore the method was the BIP detection.
Motivated by the demonstrated the feasibility, we argue to explore other forms of using the template, e.g., to compare experiences to each other.
For instance, specific questions, like the comparison of transitional environments, we suggest to present the same drawing template after each condition. 
The users then can reflect on the drawn line of the previous condition and express the latest experience by drawing the line relative to the other.
In this way, we expect more detailed insights into the temporal course of a presence experience, e.g., influenced by system design or story telling.
We found a clear tendency that the method is used to express the strength of a BIP.
Future work, therefore, might work on the elaborate the method and test its validity and reliability to transform it into a measurement tool, as questioannaires are today. 


\section{Conclusion}

In contrast to the established questionnaires -- main measurement method in more than 90\% of studies --, the usage of the drawing method gives insights into temporal effects.
Methods enabling the assessment of temporal variations, like physiological and behavioral measures, are hardly not used.

The method we provide enables users to express variations in their presence experience over time by drawing.
To enable the evaluation of the drawings, we propose a descriptive numerical model based on related works in presence research.
Our descriptive model replaces the drawings by compact numerical data for storage, comparison and evaluation purposes.

The drawing method is applied after an experience and allows the participant to draw a line that indicates the state of presence in VR over time by changing the amplitude of the line during drawing.
In an exploratory user study, we demonstrate the feasibility of the method.
We show the higher accuracy in understanding temporal variations of the presence experience compared to existing methods.
In particular, we could show that our method is feasible for detecting 100\% of breaks in the presence experience (BIP) requiring only 10 participants.
Participants self reflection on their experience was stimulated by the drawing, resulting in a very detailed expression of their presence experience.
Additionally, participants were able to express the influence of a BIP on the following cause of the experience by their line drawing.
In addition, we have good indications that, by collecting data from more participants, the method can be used to evaluate the strength of a single break in attendance.
The proposed drawing method in combination with the descriptive model closes the gap between the existing cumbersome techniques and the questionnaires that do not capture temporal variances.




\bibliographystyle{ACM-Reference-Format}
\bibliography{ms}


\begin{thebibliography}{32}


\ifx \showCODEN    \undefined \def \showCODEN     #1{\unskip}     \fi
\ifx \showDOI      \undefined \def \showDOI       #1{#1}\fi
\ifx \showISBNx    \undefined \def \showISBNx     #1{\unskip}     \fi
\ifx \showISBNxiii \undefined \def \showISBNxiii  #1{\unskip}     \fi
\ifx \showISSN     \undefined \def \showISSN      #1{\unskip}     \fi
\ifx \showLCCN     \undefined \def \showLCCN      #1{\unskip}     \fi
\ifx \shownote     \undefined \def \shownote      #1{#1}          \fi
\ifx \showarticletitle \undefined \def \showarticletitle #1{#1}   \fi
\ifx \showURL      \undefined \def \showURL       {\relax}        \fi
\providecommand\bibfield[2]{#2}
\providecommand\bibinfo[2]{#2}
\providecommand\natexlab[1]{#1}
\providecommand\showeprint[2][]{arXiv:#2}

\bibitem[\protect\citeauthoryear{Chertoff, Goldiez, and LaViola}{Chertoff
  et~al\mbox{.}}{2010}]%
        {Chertoff:2010:VET:2195920.2196260}
\bibfield{author}{\bibinfo{person}{Dustin~B. Chertoff}, \bibinfo{person}{Brian
  Goldiez}, {and} \bibinfo{person}{Joseph~J. LaViola}.}
  \bibinfo{year}{2010}\natexlab{}.
\newblock \showarticletitle{Virtual Experience Test: A Virtual Environment
  Evaluation Questionnaire}. In \bibinfo{booktitle}{\emph{Proceedings of the
  2010 IEEE Virtual Reality Conference}} \emph{(\bibinfo{series}{VR '10})}.
  \bibinfo{publisher}{IEEE Computer Society}, \bibinfo{address}{Washington, DC,
  USA}, \bibinfo{pages}{103--110}.
\newblock
\showISBNx{978-1-4244-6237-7}
\urldef\tempurl%
\url{https://doi.org/10.1109/VR.2010.5444804}
\showDOI{\tempurl}


\bibitem[\protect\citeauthoryear{Chung and Gardner}{Chung and Gardner}{2009}]%
        {Chung:2009:MTV:1670252.1670287}
\bibfield{author}{\bibinfo{person}{Jaeyong Chung} {and} \bibinfo{person}{Henry
  Gardner}.} \bibinfo{year}{2009}\natexlab{}.
\newblock \showarticletitle{Measuring Temporal Variation in Presence During
  Game Playing}. In \bibinfo{booktitle}{\emph{Proceedings of the 8th
  International Conference on Virtual Reality Continuum and Its Applications in
  Industry}} \emph{(\bibinfo{series}{VRCAI '09})}. \bibinfo{publisher}{ACM},
  \bibinfo{address}{New York, NY, USA}, \bibinfo{pages}{163--168}.
\newblock
\showISBNx{978-1-60558-912-1}
\urldef\tempurl%
\url{https://doi.org/10.1145/1670252.1670287}
\showDOI{\tempurl}


\bibitem[\protect\citeauthoryear{Chung and Gardner}{Chung and Gardner}{2012}]%
        {chung12}
\bibfield{author}{\bibinfo{person}{Jaeyong Chung} {and} \bibinfo{person}{Henry
  Gardner}.} \bibinfo{year}{2012}\natexlab{}.
\newblock \showarticletitle{Temporal Presence Variation in Immersive Computer
  Games}.
\newblock \bibinfo{journal}{\emph{International Journal of Human-Computer
  Interaction}}  \bibinfo{volume}{28} (\bibinfo{date}{08}
  \bibinfo{year}{2012}), \bibinfo{pages}{511--529}.
\newblock
\urldef\tempurl%
\url{https://doi.org/10.1080/10447318.2011.627298}
\showDOI{\tempurl}


\bibitem[\protect\citeauthoryear{Cummings and Bailenson}{Cummings and
  Bailenson}{2015}]%
        {cummings_how_2015}
\bibfield{author}{\bibinfo{person}{James Cummings} {and}
  \bibinfo{person}{Jeremy Bailenson}.} \bibinfo{year}{2015}\natexlab{}.
\newblock \showarticletitle{How Immersive Is Enough? A Meta--Analysis of the
  Effect of Immersive Technology on User Presence}.
\newblock \bibinfo{journal}{\emph{Media Psychology}} \bibinfo{volume}{19},
  \bibinfo{number}{2} (\bibinfo{year}{2015}), \bibinfo{pages}{272--309}.
\newblock
\urldef\tempurl%
\url{https://doi.org/10.1080/15213269.2015.1015740}
\showDOI{\tempurl}


\bibitem[\protect\citeauthoryear{Garau, Friedman, Widenfeld, Antley, Brogni,
  and Slater}{Garau et~al\mbox{.}}{2008}]%
        {Garau.2008}
\bibfield{author}{\bibinfo{person}{Maia Garau}, \bibinfo{person}{Doron
  Friedman}, \bibinfo{person}{Hila~Ritter Widenfeld}, \bibinfo{person}{Angus
  Antley}, \bibinfo{person}{Andrea Brogni}, {and} \bibinfo{person}{Mel
  Slater}.} \bibinfo{year}{2008}\natexlab{}.
\newblock \showarticletitle{Temporal and Spatial Variations in Presence:
  Qualitative Analysis of Interviews from an Experiment on Breaks in Presence}.
\newblock \bibinfo{journal}{\emph{Presence: Teleoper. Virtual Environ.}}
  \bibinfo{volume}{17}, \bibinfo{number}{3} (\bibinfo{year}{2008}),
  \bibinfo{pages}{293--309}.
\newblock
\showISSN{1054-7460}
\urldef\tempurl%
\url{https://doi.org/10.1162/pres.17.3.293}
\showDOI{\tempurl}


\bibitem[\protect\citeauthoryear{Heeter}{Heeter}{1992}]%
        {heeter92}
\bibfield{author}{\bibinfo{person}{Carrie Heeter}.}
  \bibinfo{year}{1992}\natexlab{}.
\newblock \showarticletitle{Being There: The Subjective Experience of
  Presence}.
\newblock \bibinfo{journal}{\emph{Presence: Teleoperators and Virtual
  Environments}} \bibinfo{volume}{1}, \bibinfo{number}{2}
  (\bibinfo{year}{1992}), \bibinfo{pages}{262--271}.
\newblock
\urldef\tempurl%
\url{https://doi.org/10.1162/pres.1992.1.2.262}
\showDOI{\tempurl}
\showeprint{https://doi.org/10.1162/pres.1992.1.2.262}


\bibitem[\protect\citeauthoryear{Hein and Mai}{Hein and Mai}{2018}]%
        {mai2018}
\bibfield{author}{\bibinfo{person}{Dimitri Hein} {and}
  \bibinfo{person}{Christian Mai}.} \bibinfo{year}{2018}\natexlab{}.
\newblock \showarticletitle{{The Usage of Presence Measurements in Research: A
  Review}}. In \bibinfo{booktitle}{\emph{Proceedings of the International
  Society for Presence Research Annual Conference}}
  \emph{(\bibinfo{series}{Presence})}. \bibinfo{publisher}{The International
  Society for Presence Research}.
\newblock


\bibitem[\protect\citeauthoryear{Ijsselsteijn and {Ridder, de}}{Ijsselsteijn
  and {Ridder, de}}{2000}]%
        {Ijsselsteijn98}
\bibfield{author}{\bibinfo{person}{Wijnand Ijsselsteijn} {and}
  \bibinfo{person}{Huib. {Ridder, de}}.} \bibinfo{year}{2000}\natexlab{}.
\newblock \showarticletitle{Measuring temporal variations in presence}. In
  \bibinfo{booktitle}{\emph{Papers from the workshop on presence in shared
  virtual environments, Ipswich, UK, June 10-11 1998}}.
\newblock


\bibitem[\protect\citeauthoryear{IJsselsteijn, {Ridder, de}, Freeman, and
  Avons}{IJsselsteijn et~al\mbox{.}}{2000}]%
        {IJsselsteijn00}
\bibfield{author}{\bibinfo{person}{Wijnand IJsselsteijn}, \bibinfo{person}{Huib
  {Ridder, de}}, \bibinfo{person}{Jonathan Freeman}, {and}
  \bibinfo{person}{Steve Avons}.} \bibinfo{year}{2000}\natexlab{}.
\newblock \showarticletitle{Presence : concept, determinants and measurement}.
  In \bibinfo{booktitle}{\emph{Human Vision and Electronic Imaging V, January
  24-27, 2000, San Jose, USA}} \emph{(\bibinfo{series}{Proceedings of SPIE -
  The International Society for Optical Engineering})},
  \bibfield{editor}{\bibinfo{person}{B.E. Rogowitz} {and} \bibinfo{person}{T.N.
  Pappas}} (Eds.). \bibinfo{publisher}{SPIE}, \bibinfo{address}{United States},
  \bibinfo{pages}{520--529}.
\newblock
\showISBNx{0-8194-3577-5}


\bibitem[\protect\citeauthoryear{Jung, Wisniewski, and Hughes}{Jung
  et~al\mbox{.}}{2018}]%
        {Sungchul18}
\bibfield{author}{\bibinfo{person}{Sungchul Jung}, \bibinfo{person}{Pamela~J.
  Wisniewski}, {and} \bibinfo{person}{Charles~E. Hughes}.}
  \bibinfo{year}{2018}\natexlab{}.
\newblock \showarticletitle{In Limbo: The Effect of Gradual Visual Transition
  Between Real and Virtual on Virtual Body Ownership Illusion and Presence}. In
  \bibinfo{booktitle}{\emph{2018 {IEEE} Conference on Virtual Reality and 3D
  User Interfaces, {VR} 2018, Tuebingen/Reutlingen, Germany, 18-22 March
  2018}}. \bibinfo{pages}{267--272}.
\newblock
\urldef\tempurl%
\url{https://doi.org/10.1109/VR.2018.8447562}
\showDOI{\tempurl}


\bibitem[\protect\citeauthoryear{Knibbe, Schjerlund, Petraeus, and
  Hornb{\ae}k}{Knibbe et~al\mbox{.}}{2018}]%
        {Knibbe:2018:DCE:3173574.3174057}
\bibfield{author}{\bibinfo{person}{Jarrod Knibbe}, \bibinfo{person}{Jonas
  Schjerlund}, \bibinfo{person}{Mathias Petraeus}, {and}
  \bibinfo{person}{Kasper Hornb{\ae}k}.} \bibinfo{year}{2018}\natexlab{}.
\newblock \showarticletitle{The Dream is Collapsing: The Experience of Exiting
  VR}. In \bibinfo{booktitle}{\emph{Proceedings of the 2018 CHI Conference on
  Human Factors in Computing Systems}} \emph{(\bibinfo{series}{CHI '18})}.
  \bibinfo{publisher}{ACM}, \bibinfo{address}{New York, NY, USA}, Article
  \bibinfo{articleno}{483}, \bibinfo{numpages}{13}~pages.
\newblock
\showISBNx{978-1-4503-5620-6}
\urldef\tempurl%
\url{https://doi.org/10.1145/3173574.3174057}
\showDOI{\tempurl}


\bibitem[\protect\citeauthoryear{Kohli}{Kohli}{2013}]%
        {Kohli:2013:RT:2519692}
\bibfield{author}{\bibinfo{person}{Luv Kohli}.}
  \bibinfo{year}{2013}\natexlab{}.
\newblock \emph{\bibinfo{title}{Redirected Touching}}.
\newblock \bibinfo{thesistype}{Ph.D. Dissertation}. \bibinfo{address}{Chapel
  Hill, NC, USA}.
\newblock Advisor(s) Brooks,Jr., Frederick P.
\newblock
\showISBNx{978-1-303-10559-3}
\newblock
\shownote{AAI3562754.}


\bibitem[\protect\citeauthoryear{Kujala, Roto,
  V\"{a}\"{a}n\"{a}nen-Vainio-Mattila, Karapanos, and Sinnel\"{a}}{Kujala
  et~al\mbox{.}}{2011}]%
        {Kujala:2011:UCM:2051372.2051531}
\bibfield{author}{\bibinfo{person}{Sari Kujala}, \bibinfo{person}{Virpi Roto},
  \bibinfo{person}{Kaisa V\"{a}\"{a}n\"{a}nen-Vainio-Mattila},
  \bibinfo{person}{Evangelos Karapanos}, {and} \bibinfo{person}{Arto
  Sinnel\"{a}}.} \bibinfo{year}{2011}\natexlab{}.
\newblock \showarticletitle{UX Curve: A Method for Evaluating Long-term User
  Experience}.
\newblock \bibinfo{journal}{\emph{Interact. Comput.}} \bibinfo{volume}{23},
  \bibinfo{number}{5} (\bibinfo{date}{Sept.} \bibinfo{year}{2011}),
  \bibinfo{pages}{473--483}.
\newblock
\showISSN{0953-5438}
\urldef\tempurl%
\url{https://doi.org/10.1016/j.intcom.2011.06.005}
\showDOI{\tempurl}


\bibitem[\protect\citeauthoryear{Liebold, Brill, Pietschmann, Schwab, and
  Ohler}{Liebold et~al\mbox{.}}{2017}]%
        {Liebold2017}
\bibfield{author}{\bibinfo{person}{Benny Liebold}, \bibinfo{person}{Michael
  Brill}, \bibinfo{person}{Daniel Pietschmann}, \bibinfo{person}{Frank Schwab},
  {and} \bibinfo{person}{Peter Ohler}.} \bibinfo{year}{2017}\natexlab{}.
\newblock \showarticletitle{Continuous Measurement of Breaks in Presence:
  Psychophysiology and Orienting Responses}.
\newblock \bibinfo{journal}{\emph{Media Psychology}} \bibinfo{volume}{20},
  \bibinfo{number}{3} (\bibinfo{year}{2017}), \bibinfo{pages}{477--501}.
\newblock
\urldef\tempurl%
\url{https://doi.org/10.1080/15213269.2016.1206829}
\showDOI{\tempurl}
\showeprint{https://doi.org/10.1080/15213269.2016.1206829}


\bibitem[\protect\citeauthoryear{Mai and Khamis}{Mai and Khamis}{2018}]%
        {mai2018perdis}
\bibfield{author}{\bibinfo{person}{Christian Mai} {and}
  \bibinfo{person}{Mohamed Khamis}.} \bibinfo{year}{2018}\natexlab{}.
\newblock \showarticletitle{{Public HMDs: Modeling and Understanding User
  Behavior around Public Head-Mounted Displays}}. In
  \bibinfo{booktitle}{\emph{Proceedings of the 7th ACM International Symposium
  on Pervasive Displays}} \emph{(\bibinfo{series}{PerDis '18})}.
  \bibinfo{publisher}{ACM}, \bibinfo{address}{New York, NY, USA}.
\newblock
\urldef\tempurl%
\url{https://doi.org/10.1145/3205873.3205879}
\showDOI{\tempurl}


\bibitem[\protect\citeauthoryear{Mai, Wiltzius, Alt, and Hu{\ss}mann}{Mai
  et~al\mbox{.}}{2018}]%
        {Mai_}
\bibfield{author}{\bibinfo{person}{Christian Mai}, \bibinfo{person}{Tim
  Wiltzius}, \bibinfo{person}{Florian Alt}, {and} \bibinfo{person}{Heinrich
  Hu{\ss}mann}.} \bibinfo{year}{2018}\natexlab{}.
\newblock \showarticletitle{{Feeling Alone in Public. Investigating the
  Influence of Spatial Layout on Users' VR Experience}}. In
  \bibinfo{booktitle}{\emph{Proceedings of the 10th Nordic Conference on
  Human-Computer Interaction}} \emph{(\bibinfo{series}{NordiCHI '18})}.
  \bibinfo{publisher}{ACM}, \bibinfo{address}{New York, NY, USA}.
\newblock
\urldef\tempurl%
\url{https://doi.org/10.1145/3240167.3240200}
\showDOI{\tempurl}


\bibitem[\protect\citeauthoryear{Minsky}{Minsky}{1980}]%
        {minsky_telepresence_1980}
\bibfield{author}{\bibinfo{person}{Marvin Minsky}.}
  \bibinfo{year}{1980}\natexlab{}.
\newblock \showarticletitle{Telepresence}.
\newblock \bibinfo{journal}{\emph{Omni}} \bibinfo{volume}{2},
  \bibinfo{number}{9} (\bibinfo{year}{1980}), \bibinfo{pages}{45--52}.
\newblock


\bibitem[\protect\citeauthoryear{Nielsen and Landauer}{Nielsen and
  Landauer}{1993}]%
        {Nielsen:1993:MMF:169059.169166}
\bibfield{author}{\bibinfo{person}{Jakob Nielsen} {and}
  \bibinfo{person}{Thomas~K. Landauer}.} \bibinfo{year}{1993}\natexlab{}.
\newblock \showarticletitle{A Mathematical Model of the Finding of Usability
  Problems}. In \bibinfo{booktitle}{\emph{Proceedings of the INTERACT '93 and
  CHI '93 Conference on Human Factors in Computing Systems}}
  \emph{(\bibinfo{series}{CHI '93})}. \bibinfo{publisher}{ACM},
  \bibinfo{address}{New York, NY, USA}, \bibinfo{pages}{206--213}.
\newblock
\showISBNx{0-89791-575-5}
\urldef\tempurl%
\url{https://doi.org/10.1145/169059.169166}
\showDOI{\tempurl}


\bibitem[\protect\citeauthoryear{Oberd{\"{o}}rfer, Fischbach, and
  Latoschik}{Oberd{\"{o}}rfer et~al\mbox{.}}{2018}]%
        {OberdorferFL18}
\bibfield{author}{\bibinfo{person}{Sebastian Oberd{\"{o}}rfer},
  \bibinfo{person}{Martin Fischbach}, {and} \bibinfo{person}{Marc~Erich
  Latoschik}.} \bibinfo{year}{2018}\natexlab{}.
\newblock \showarticletitle{Effects of {VE} Transition Techniques on Presence,
  Illusion of Virtual Body Ownership, Efficiency, and Naturalness}. In
  \bibinfo{booktitle}{\emph{Proceedings of the Symposium on Spatial User
  Interaction, {SUI} 2018, Berlin, Germany, October 13-14, 2018}}.
  \bibinfo{pages}{89--99}.
\newblock
\urldef\tempurl%
\url{https://doi.org/10.1145/3267782.3267787}
\showDOI{\tempurl}


\bibitem[\protect\citeauthoryear{Rosakranse and Oh}{Rosakranse and Oh}{2014}]%
        {rosa.14}
\bibfield{author}{\bibinfo{person}{Christine Rosakranse} {and}
  \bibinfo{person}{Soo~Youn Oh}.} \bibinfo{year}{2014}\natexlab{}.
\newblock \showarticletitle{Measuring Presence: The Use Trends of Five
  Canonical Presence Questionaires from 1998--2012}. In
  \bibinfo{booktitle}{\emph{In Proceedings of the 15th International Workshop
  on Presence (ISPR’14)}}.
\newblock


\bibitem[\protect\citeauthoryear{Sahar, Varsaluoma, and Kujala}{Sahar
  et~al\mbox{.}}{2014}]%
        {Sahar:2014:CEE:2676467.2676500}
\bibfield{author}{\bibinfo{person}{Farrukh Sahar}, \bibinfo{person}{Jari
  Varsaluoma}, {and} \bibinfo{person}{Sari Kujala}.}
  \bibinfo{year}{2014}\natexlab{}.
\newblock \showarticletitle{Comparing the Effectiveness of Electronic Diary and
  UX Curve Methods in Multi-component Product Study}. In
  \bibinfo{booktitle}{\emph{Proceedings of the 18th International Academic
  MindTrek Conference: Media Business, Management, Content \& Services}}
  \emph{(\bibinfo{series}{AcademicMindTrek '14})}. \bibinfo{publisher}{ACM},
  \bibinfo{address}{New York, NY, USA}, \bibinfo{pages}{93--100}.
\newblock
\showISBNx{978-1-4503-3006-0}
\urldef\tempurl%
\url{https://doi.org/10.1145/2676467.2676500}
\showDOI{\tempurl}


\bibitem[\protect\citeauthoryear{Schubert}{Schubert}{2003}]%
        {Schubert2003}
\bibfield{author}{\bibinfo{person}{Thomas Schubert}.}
  \bibinfo{year}{2003}\natexlab{}.
\newblock \showarticletitle{The sense of presence in virtual environments: A
  three-component scale measuring spatial presence, involvement, and realness.}
\newblock \bibinfo{journal}{\emph{Zeitschrift f{\"u}r Medienpsychologie}}
  \bibinfo{volume}{15}, \bibinfo{number}{2} (\bibinfo{year}{2003}),
  \bibinfo{pages}{69--71}.
\newblock
\urldef\tempurl%
\url{https://doi.org/10.1026//1617-6383.15.2.69}
\showDOI{\tempurl}


\bibitem[\protect\citeauthoryear{Schubert, Friedmann, and Regenbrecht}{Schubert
  et~al\mbox{.}}{2001}]%
        {schubert_experience_2001}
\bibfield{author}{\bibinfo{person}{Thomas Schubert}, \bibinfo{person}{Frank
  Friedmann}, {and} \bibinfo{person}{Holger Regenbrecht}.}
  \bibinfo{year}{2001}\natexlab{}.
\newblock \showarticletitle{The {Experience} of {Presence}: {Factor} {Analytic}
  {Insights}}.
\newblock \bibinfo{journal}{\emph{Presence: Teleoperators and Virtual
  Environments}} \bibinfo{volume}{10}, \bibinfo{number}{3}
  (\bibinfo{date}{June} \bibinfo{year}{2001}), \bibinfo{pages}{266--281}.
\newblock
\showISSN{1054-7460, 1531-3263}
\urldef\tempurl%
\url{https://doi.org/10.1162/105474601300343603}
\showDOI{\tempurl}


\bibitem[\protect\citeauthoryear{Slater}{Slater}{2009a}]%
        {Slater3549}
\bibfield{author}{\bibinfo{person}{Mel Slater}.}
  \bibinfo{year}{2009}\natexlab{a}.
\newblock \showarticletitle{Place illusion and plausibility can lead to
  realistic behaviour in immersive virtual environments}.
\newblock \bibinfo{journal}{\emph{Philosophical Transactions of the Royal
  Society of London B: Biological Sciences}} \bibinfo{volume}{364},
  \bibinfo{number}{1535} (\bibinfo{year}{2009}), \bibinfo{pages}{3549--3557}.
\newblock
\showISSN{0962-8436}
\urldef\tempurl%
\url{https://doi.org/10.1098/rstb.2009.0138}
\showDOI{\tempurl}
\showeprint{http://rstb.royalsocietypublishing.org/content/364/1535/3549.full.pdf}


\bibitem[\protect\citeauthoryear{Slater}{Slater}{2009b}]%
        {Slater.2009}
\bibfield{author}{\bibinfo{person}{Mel Slater}.}
  \bibinfo{year}{2009}\natexlab{b}.
\newblock \showarticletitle{Place illusion and plausibility can lead to
  realistic behaviour in immersive virtual environments}.
\newblock \bibinfo{journal}{\emph{Philosophical Transactions of the Royal
  Society of London B: Biological Sciences}} \bibinfo{volume}{364},
  \bibinfo{number}{1535} (\bibinfo{year}{2009}), \bibinfo{pages}{3549--3557}.
\newblock
\showISSN{0962-8436}
\urldef\tempurl%
\url{https://doi.org/10.1098/rstb.2009.0138}
\showDOI{\tempurl}
\showeprint{http://rstb.royalsocietypublishing.org/content/364/1535/3549.full.pdf}


\bibitem[\protect\citeauthoryear{Slater and Steed}{Slater and Steed}{2000}]%
        {Slater:2000}
\bibfield{author}{\bibinfo{person}{Mel Slater} {and} \bibinfo{person}{Anthony
  Steed}.} \bibinfo{year}{2000}\natexlab{}.
\newblock \showarticletitle{A Virtual Presence Counter}.
\newblock \bibinfo{journal}{\emph{Presence: Teleoperators and Virtual
  Environments}} \bibinfo{volume}{9}, \bibinfo{number}{5} (\bibinfo{date}{Oct.}
  \bibinfo{year}{2000}), \bibinfo{pages}{413--434}.
\newblock
\showISSN{1054-7460}
\urldef\tempurl%
\url{https://doi.org/10.1162/105474600566925}
\showDOI{\tempurl}


\bibitem[\protect\citeauthoryear{Slater and Wilbur}{Slater and Wilbur}{1997}]%
        {Slater97}
\bibfield{author}{\bibinfo{person}{Mel Slater} {and} \bibinfo{person}{Sylvia
  Wilbur}.} \bibinfo{year}{1997}\natexlab{}.
\newblock \showarticletitle{A Framework for Immersive Virtual Environments
  (FIVE): Speculations on the Role of Presence in Virtual Environments}.
\newblock \bibinfo{journal}{\emph{Presence: Teleoperators and Virtual
  Environments}} \bibinfo{volume}{6}, \bibinfo{number}{6}
  (\bibinfo{year}{1997}), \bibinfo{pages}{603--616}.
\newblock
\urldef\tempurl%
\url{https://doi.org/10.1162/pres.1997.6.6.603}
\showDOI{\tempurl}
\showeprint{https://doi.org/10.1162/pres.1997.6.6.603}


\bibitem[\protect\citeauthoryear{Sproll, Freiberg, Grechkin, and Riecke}{Sproll
  et~al\mbox{.}}{2013}]%
        {sproll13}
\bibfield{author}{\bibinfo{person}{Daniel Sproll}, \bibinfo{person}{Jacob
  Freiberg}, \bibinfo{person}{Timofey Grechkin}, {and}
  \bibinfo{person}{Bernhard Riecke}.} \bibinfo{year}{2013}\natexlab{}.
\newblock \showarticletitle{Poster: Paving the way into virtual reality - A
  transition in five stages}. In \bibinfo{booktitle}{\emph{IEEE Symposium on 3D
  User Interface 2013 - Proceedings}} \emph{(\bibinfo{series}{SUI '13})}.
  \bibinfo{pages}{175--176}.
\newblock
\showISBNx{978-1-4673-6097-5}


\bibitem[\protect\citeauthoryear{Steinicke, Bruder, Hinrichs, Steed, and
  Gerlach}{Steinicke et~al\mbox{.}}{2009}]%
        {4811024}
\bibfield{author}{\bibinfo{person}{Frank Steinicke}, \bibinfo{person}{Gerd
  Bruder}, \bibinfo{person}{Klaus Hinrichs}, \bibinfo{person}{Anthony Steed},
  {and} \bibinfo{person}{Alexander Gerlach}.} \bibinfo{year}{2009}\natexlab{}.
\newblock \showarticletitle{Does a Gradual Transition to the Virtual World
  increase Presence?}. In \bibinfo{booktitle}{\emph{2009 IEEE Virtual Reality
  Conference}}. \bibinfo{pages}{203--210}.
\newblock
\showISSN{1087-8270}
\urldef\tempurl%
\url{https://doi.org/10.1109/VR.2009.4811024}
\showDOI{\tempurl}


\bibitem[\protect\citeauthoryear{Valkov and Flagge}{Valkov and Flagge}{2017}]%
        {Valkov:2017:SIB:3131277.3132183}
\bibfield{author}{\bibinfo{person}{Dimitar Valkov} {and}
  \bibinfo{person}{Steffen Flagge}.} \bibinfo{year}{2017}\natexlab{}.
\newblock \showarticletitle{Smooth Immersion: The Benefits of Making the
  Transition to Virtual Environments a Continuous Process}. In
  \bibinfo{booktitle}{\emph{Proceedings of the 5th Symposium on Spatial User
  Interaction}} \emph{(\bibinfo{series}{SUI '17})}. \bibinfo{publisher}{ACM},
  \bibinfo{address}{New York, NY, USA}, \bibinfo{pages}{12--19}.
\newblock
\showISBNx{978-1-4503-5486-8}
\urldef\tempurl%
\url{https://doi.org/10.1145/3131277.3132183}
\showDOI{\tempurl}


\bibitem[\protect\citeauthoryear{Varsaluoma and Sahar}{Varsaluoma and
  Sahar}{2014}]%
        {Varsaluoma:2014:MRU:2676467.2676485}
\bibfield{author}{\bibinfo{person}{Jari Varsaluoma} {and}
  \bibinfo{person}{Farrukh Sahar}.} \bibinfo{year}{2014}\natexlab{}.
\newblock \showarticletitle{Measuring Retrospective User Experience of
  Non-powered Hand Tools: An Exploratory Remote Study with UX Curve}. In
  \bibinfo{booktitle}{\emph{Proceedings of the 18th International Academic
  MindTrek Conference: Media Business, Management, Content \& Services}}
  \emph{(\bibinfo{series}{AcademicMindTrek '14})}. \bibinfo{publisher}{ACM},
  \bibinfo{address}{New York, NY, USA}, \bibinfo{pages}{40--47}.
\newblock
\showISBNx{978-1-4503-3006-0}
\urldef\tempurl%
\url{https://doi.org/10.1145/2676467.2676485}
\showDOI{\tempurl}


\bibitem[\protect\citeauthoryear{Witmer and Singer}{Witmer and Singer}{1998}]%
        {Witmer98}
\bibfield{author}{\bibinfo{person}{Bob~G. Witmer} {and}
  \bibinfo{person}{Michael~J. Singer}.} \bibinfo{year}{1998}\natexlab{}.
\newblock \showarticletitle{Measuring Presence in Virtual Environments: A
  Presence Questionnaire}.
\newblock \bibinfo{journal}{\emph{Presence: Teleoper. Virtual Environ.}}
  \bibinfo{volume}{7}, \bibinfo{number}{3} (\bibinfo{date}{June}
  \bibinfo{year}{1998}), \bibinfo{pages}{225--240}.
\newblock
\showISSN{1054-7460}
\urldef\tempurl%
\url{https://doi.org/10.1162/105474698565686}
\showDOI{\tempurl}


\end{thebibliography}

\end{document}